\documentclass[prb,twocolumn,showpacs,amsmath,amssymb,citeautoscript]{revtex4}
\usepackage{graphicx}
\usepackage{dcolumn}
\usepackage{amssymb}
\usepackage{amsmath}
\usepackage{braket}
\usepackage{verbatim}

\begin{document}

\title{First-principles theory of nonradiative carrier capture via multiphonon emission}

\author{Audrius Alkauskas}
\author{Qimin Yan}
\author{Chris G. Van de Walle}

\affiliation{Materials Department, University of California, Santa Barbara, California 93106-5050, USA}
\date{\today}

\begin{abstract}
We develop a practical first-principles methodology to determine nonradiative carrier capture
coefficients at defects in semiconductors. We consider transitions that occur via multiphonon emission.
Parameters in the theory, including electron-phonon coupling matrix elements, are computed consistently
using state-of-the-art electronic structure techniques based on hybrid density functional theory. These provide
a significantly improved description of bulk band structures, as well as defect geometries and wavefunctions.
In order to properly describe carrier capture processes at charged centers, we put forward an approach to treat
the effect of long-range Coulomb interactions on scattering states in the framework of supercell calculations.
We also discuss the choice of initial conditions for a perturbative treatment of carrier capture. As a benchmark,
we apply our theory to several hole-capturing centers in GaN and ZnO, materials of high technological importance
in which the role of defects is being actively investigated.  Calculated hole capture coefficients are in
good agreement with experimental data. We discuss the insights gained into the physics of defects in wide-band-gap
semiconductors, such as the strength of electron-phonon coupling and the role of different phonon modes.

\end{abstract}

\maketitle


\section{Introduction}
\label{intro}

Point defects drastically affect the performance of semiconductor devices. In particular, they can
act as charge traps and/or recombination centers. In electronic applications, such as in high-electron
mobility transistors, charge traps deteriorate the performance of the device and can lead to so-called
device dispersion.\cite{Mishra_IEEE_2002} In most cases charge trapping, or capture, occurs nonradiatively,
i.e., without the emission of a photon. In optoelectronic applications, such as in light-emitting diodes
or photovoltaic cells, defects can act as recombination centers for charge carriers. This so-called
Shockley-Read-Hall (SRH) recombination \cite{Abakumov} is detrimental, as it decreases the efficiency
of the device. SRH recombination can also affect electronic devices that rely on minority carrier transport,
e.g., bipolar transistors.  SRH recombination is a sequence  of two carrier capture processes: one carrier
is captured, and then the other carrier recombines with it. \cite{Abakumov} For both charge traps and
recombination centers, the important question is: what are the carrier capture coefficients (cross
sections)?

For deep centers the nonradiative carrier capture occurs via multiphonon emission (MPE).
\cite{Abakumov,DiBartolo,Henry_PRB_1977} The main idea behind MPE is that the transition between
the delocalized bulk state and the localized defect state can occur within the first order of
electron-phonon coupling because of a large local lattice relaxation associated with the change
of the charge state of the defect. \cite{Abakumov,DiBartolo,Henry_PRB_1977} The phonon selection
rule $\Delta n = \pm 1$ is relieved, and emission of more than one phonon becomes possible. Many
researchers have contributed to the theoretical foundations of MPE over the past six decades.
\cite{Huang_PRCL_1950,Kubo_PTP_1955,Gummel_AP_1957,Henry_PRB_1977,Freed_JCP_1970,Paessler_pssb_1975,
Huang_SS_1981,Gutsche_pssb_1982,Peuker_pssb_1982,Burt_JPC_1983,Paessler_CJP_1982,Goguenheim_JAP_1990}
These investigations have revealed that the results of calculations are extremely sensitive to (i) the
adopted theoretical model and (ii) the details of the electronic structure of the defect, with different
approaches yielding variations of capture coefficients over many orders of magnitude. \cite{Stoneham}

Concerning aspect (i), earlier theoretical works \cite{Huang_PRCL_1950,Kubo_PTP_1955,Gummel_AP_1957,
Henry_PRB_1977,Freed_JCP_1970,Paessler_pssb_1975,Huang_SS_1981,Gutsche_pssb_1982,Peuker_pssb_1982,
Burt_JPC_1983,Paessler_CJP_1982,Goguenheim_JAP_1990} have made it clear that there is no single
theoretical model that is valid in {\it all} cases. A number of factors have to be considered
in choosing the appropriate description, \cite{Huang_PRCL_1950,Kubo_PTP_1955,Gummel_AP_1957,
Henry_PRB_1977,Freed_JCP_1970,Paessler_pssb_1975,Huang_SS_1981,Gutsche_pssb_1982,Peuker_pssb_1982,
Burt_JPC_1983,Paessler_CJP_1982,Goguenheim_JAP_1990} including the hierarchy of different time
scales (carrier capture times vs.\ phonon lifetimes and periods of lattice vibrations), the
strength of electron-phonon coupling (linear vs.\ higher-order coupling schemes), the choice
of a good starting point for perturbation theory (electron and phonon wavefunctions), and
the number of different phonon modes that have to be considered. This choice of description
has to be considered for each type of defect individually, a practice we will follow in the
current paper as well.

Aspect (ii), i.e., incomplete knowledge of the atomic and electronic structure of the defect, turned
out to be an equally important issue. If this structure is not known, not only does it affect the result
within a given theoretical model, but it impedes the choice of the correct model itself. Aspect (ii)
is thus closely linked to aspect (i). When the objective was to understand general trends and interpret
experimental findings, calculations based on models that did not take the specifics of the atomic and
the electronic structure into account were often very successful. An example of such work is the
seminal paper of Henry and Lang \cite{Henry_PRB_1977} on nonradiative carrier capture in GaP and GaAs,
semiconductors with room-temperature band gaps of 2.22 and 1.42 eV,\cite{Strehlow_JPCRD_1973} respectively.
The authors theoretically determined the temperature dependence of capture cross sections and provided an
estimate of the range of high-temperature asymptotic values of these cross sections. Using a semi-classical
description of carrier capture, they could explain the exponential dependence of cross sections on temperature
for many defects in both GaAs and GaP, which proved that for these systems carrier capture was indeed due
to multiphonon emission. However, their model was unable to offer specific predictions for individual defects,
and provided little insight into exceptions to the general trends. In addition, these as well as other
early calculations required empirical input as well as drastic simplifications regarding the local electronic
structure and the nature of relevant lattice vibrations. This seriously limited the predictive power, especially
for applications to new materials.

With the advent of accurate electronic structure methods, mostly based on density functional theory (DFT)
and related techniques, the situation is very different now. \cite{VanDeWalle_JAP_2004, Estreicher}
State-of-the-art approaches, such as hybrid functionals, provide a very good description of both bulk band
gaps and localized defect states. \cite{Wiley} The availability of these methods, combined with the general
knowledge of MPE acquired over the past six decades, raises the question whether nonradiative carrier capture
rates can now be determined completely from first principles, allowing them to be used predictively, and
whether such calculations can expand our insights into the \emph{physics} of defects. This provides the
motivation for our work.

Some progress has already been made in this area. Schanovsky and co-workers studied nonradiative hole trapping
at defects in SiO$_2$ and addressed the vibrational part of the problem using first-principles calculations,
\cite{Schanovsky_JVST_2011,Schanovsky_JCE_2012} but actual values of the electron-phonon matrix elements remained undetermined.
McKenna and Blumberger \cite{McKenna_PRB_2012} studied the related problem of electron transfer between defect
states within the Marcus theory, \cite{Marcus_RMP_1993} and determined the electron-phonon coupling matrix element
between two localized defect states in MgO directly from electronic structure calculations. Shi and Wang
\cite{Shi_PRL_2012} were the first to address both the vibrational and the electron-phonon part of the carrier-capture
problem completely from electronic structure calculations. They presented an algorithm to calculate electron-phonon
matrix elements at defects, and applied the methodology to study hole capture at the Zn$_{\text{Ga}}$-$V_{\text{N}}$
complex in GaN. Despite some important contributions, this study also had some limitations. First, the theory was
applied to a defect for which direct experimental data is not available. \cite{Reshchikov_PRB_2011} Second, both
ground-state geometries of the defect and electron-phonon matrix elements were determined using a semilocal
functional within the so-called generalized gradient approximation (GGA). Such functionals underestimate bulk
band gaps and tend to over-delocalize defect wavefunctions.  As discussed in Sec.\ \ref{meth}, more accurate
approaches are available that overcome these drawbacks.
Third, as we analyze in Sec.\ \ref{disc}, the theoretical approach used in Ref.~ \onlinecite{Shi_PRL_2012},
the so-called adiabatic formulation within the Condon approximation, can be questioned for
describing nonradiative capture at defects. \cite{Huang_SS_1981,Peuker_pssb_1982}

Overall, it is clear that the current status of modeling nonradiative capture at defects in solids is still
unsatisfactory, especially when contrasted with the impressive advances in treating electron-phonon coupling in
defect-free crystals, \cite{Baroni_RMP_2001,Giustino_PRB_2007} or in describing nonradiative processes in
molecules.\cite{Borrelli_JCP_2003}

In this work we present calculations of carrier capture rates via MPE entirely from first principles. The
electronic structure, the vibrational properties, and the electron-phonon coupling are determined from
accurate electronic structure techniques, in particular hybrid density functional theory. Specifically, we
present a method to calculate electron-phonon coupling matrix elements at defects consistently within the
hybrid functional approach. Our calculations yield absolute carrier capture rates without any fitting parameters.
We apply the methodology to a set of defects in GaN and ZnO, wide-band-gap semiconductors with $T$=$0$ K band
gaps of 3.50 \cite{Nepal_APL_2005} and 3.44 eV,\cite{Reynolds_PRB_1999} respectively. We first study
C$_{\text{N}}$ in GaN and Li$_{\text{Zn}}$ in ZnO because optical signatures of these two defects
are well established \cite{Reshchikov_JAP_2005,Reshchikov_PB_2007,Reshchikov_MRS_2007,Reshchikov_AIP_2014,Ogino_JJAP_1980,
Meyer_pss_2004,Lyons_APL_2009} and nonradiative capture coefficients are available.\cite{Reshchikov_AIP_2014}
We also apply our methodology to the Zn$_{\text{Ga}}$-$V_{\text{N}}$ defect in GaN to compare our results
with those of Ref.~ \onlinecite{Shi_PRL_2012}.

This paper is organized as follows. The problem of nonradiative carrier capture is described in Section \ref{def}.
In Section \ref{meth} we outline the theoretical formulation of the MPE, present technical details of our computational
toolbox, and discuss how various quantities are calculated. In Section \ref{res} we present results for selected defects
in GaN and ZnO and compare with available experimental data and other computational approaches. In Section \ref{disc}
we critically analyze our approach and discuss insights gained into defect physics in GaN and ZnO.
Section \ref{conc} concludes the paper.


\begin{figure}
\includegraphics[width=8.5cm]{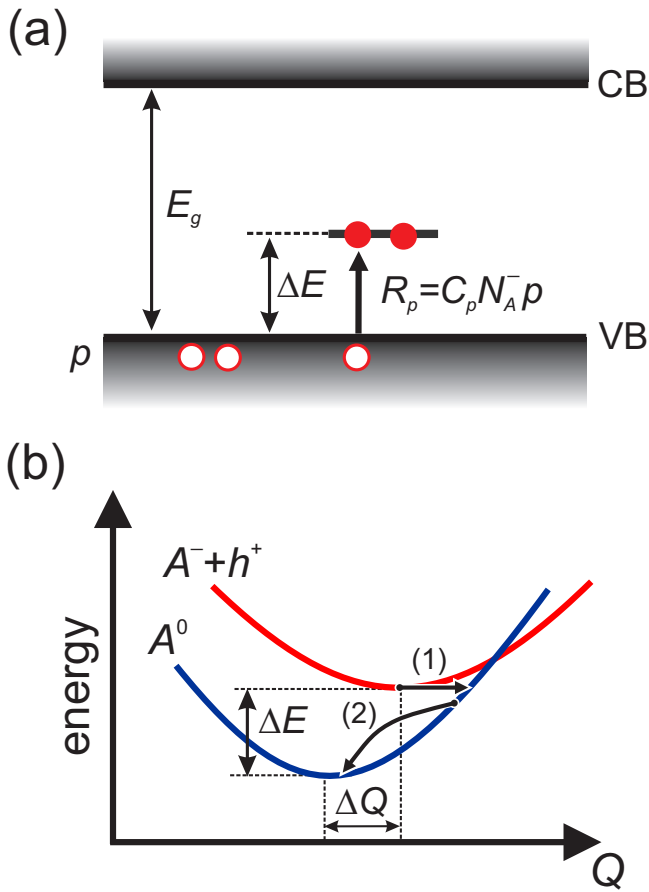}
\caption{(Color online) Nonradiative carrier capture at a deep defect in two representations: (a) band diagram
and (b) configuration coordinate diagram. For illustration purposes, the defect is a deep acceptor with a
negative (doubly-occupied) and a neutral (singly-occupied) charge state.  $\Delta E$ is the ionization energy
of the acceptor, and $Q$ is an appropriately chosen configuration coordinate. In (b), process (1) is the change
of the electronic state due to electron-phonon coupling, process (2) is vibrational relaxation due to phonon-phonon
interactions.
}
\label{SRH}
\end{figure}

\section{Definition of the problem \label{def}}

Without loss of generality, let us consider nonradiative carrier capture of a hole by an
acceptor defect. The process is illustrated in Fig.\ \ref{SRH} in two different representations:
(a) a band diagram, and (b) a configuration coordinate (cc) diagram. In the latter, a
one-dimensional generalized coordinate $Q$ is used to represent atomic relaxations. \cite{Stoneham}
The excited state of the system corresponds to the negatively charged acceptor and
a hole in the valence band ($A^{-}$+$h^{+}$), while the ground state corresponds to
the neutral state of the acceptor ($A^{0}$). The equilibrium geometries of the two
charge states are different. $\Delta E$ is the energy difference between the two states.

Carrier capture consists of two elementary steps: an energy-conserving transition between
two electronic states, process (1), and vibrational relaxation, process (2), in Fig.\ \ref{SRH}(b).
Vibrational relaxation occurs on a timescale of a few picoseconds,\cite{Matulionis_APL_2009}
while the electronic transition is much slower.\cite{Henry_PRB_1977} Thus, the electronic transition
is the bottleneck for nonradiative capture, and in this work we will only consider process (1).

Let $p$ be the density of holes in the system. The total concentration of defects is $N_A=N_A^{0}+N_A^{-}$,
where $N_A^{-}$ is the density of negatively charged (ionized) acceptors, and $N_A^{0}$ is the density of
neutral acceptors. The holes are captured at a rate \cite{Abakumov}
\begin{equation}
R_{p}=C_{p}N_A^{-}p,
\label{rate}
\end{equation}
where the units of $R_{p}$ are cm$^{-3}$s$^{-1}$; $C_{p}$ is the  hole capture coefficient, with units
$[C_{p}]$=cm$^3$s$^{-1}$.
An analogous equation applies to electron capture processes.

In principle carrier capture can occur both radiatively and nonradiatively. \cite{Huang_PRCL_1950}
The two processes are in general competing and can occur simultaneously. The rate of radiative transitions
increases with the energy of a transition as a power law; for semiconductors typical capture coefficients
are of the order $C_{\{n,p\}}\sim10^{-14}-10^{-13}$ cm$^{3}$s$^{-1}$. \cite{Stoneham} The dependence of
nonradiative capture rates on the energy of the transition $\Delta E$ is usually nonmonotonic;
capture coefficients can vary over a very wide range $C_{\{n,p\}}\sim10^{-14}-10^{-6}$ cm$^{3}$s$^{-1}$.
\cite{Henry_PRB_1977,Stoneham} When capture coefficients are in the upper part of this range, nonradiative
transitions are dominant, and radiative transitions can be neglected. This is the case for all capture
processes that we study in the present work.

The main goal of the theory is to determine nonradiative electron and hole capture coefficients $C_n$ and
$C_p$ from electronic structure calculations. In the literature, carrier capture processes are often described
in terms of capture cross sections $\sigma$. The two quantities are related via  $C=\left\langle v\right\rangle\sigma$,
where $\left\langle v\right\rangle$ is a characteristic electron velocity. For non-degenerate statistics
this velocity is the average thermal velocity. While $C_n$ and $C_p$ are more fundamental quantities,
capture cross sections are useful because of their straightforward and intuitive interpretation. Experimental
values for capture cross sections in a wide variety of systems \cite{Henry_PRB_1977,Stoneham} vary between
$10^{-5}$ \AA $^2$ (weak coupling) and $\sim 10^3$ \AA $^{2}$ (very strong coupling).


\section{Theoretical formulation and computational methodology \label{meth}}

\subsection{Computational toolbox \label{Comp}}

To describe the atomic and the electronic structure of defects and bulk materials we use
DFT with a hybrid functional. \cite{Heyd_JCP_2003} Hybrid functionals add a fraction $\alpha$ of
Fock exchange to the exchange described by the generalized gradient approximation, greatly
improving the description of structural properties and band structures, including band gaps.
Both of these aspects are particularly important for defects.\cite{Pacchioni_PRB_2000,Deak_JPCM_2005,
Alkauskas_PRL_2008,Lyons_APL_2009} In addition, hybrid functionals can correctly describe the
polaronic nature of anion-bound holes derived from N and O $2p$ states, \cite{Pacchioni_PRB_2000,
Lyons_APL_2009,Lyons_PRL_2012,Lyons_JJAP_2013} which is crucial for the defects in the present study.

We use the functional of Heyd, Scuseria, and Ernzerhof (HSE). \cite{Heyd_JCP_2003}
In this functional the Fock exchange is screened (screening parameter $\mu$=$0.2$ {\AA}$^{-1}$),
and the sum rule for the exchange hole is fulfilled by suitably modifying the semilocal part of the
exchange. We adapt the functional by tuning $\alpha$ to reproduce the experimental band gaps, which
has become a common procedure; \cite{Deak_JPCM_2005,Alkauskas_PRL_2008,Lyons_APL_2009}
the corresponding values are $\alpha$=$0.31$ for GaN and $\alpha$=$0.38$ for ZnO. For $\alpha$=$0$ and
$\mu$=$0$ the HSE functional does not contain nonlocal exchange and is identical to the generalized
gradient approximation functional of Perdew, Burke, and Ernzerhof (PBE). \cite{Perdew_PRL_1996}

Our electronic structure calculations are based on the projector-augmented wave (PAW) formalism
\cite{Bloechl_PRB_1994}, with PAW potentials generated at the PBE level. We have used the {\sc vasp}
code \cite{VASP} with the implementation of hybrid functionals described in Ref.~\ \onlinecite{Paier_JCP_2006}.
A kinetic energy cutoff of 29.4 Ry (400 eV) was used in all calculations.
In the case of Zn, $3d$ states were included in the valence.
The resulting lattice parameters are  $a$=$3.20$ \AA,
$c$=$5.19$ \AA, and $u$=$0.377$ for GaN (in excellent agreement with the experimental \cite{Madelung}
values 3.19 \AA, 5.20 \AA, and 0.377, respectively); and $a$=$3.24$ \AA, $c$=$5.21$ \AA,
and $u$=$0.379$ for ZnO (experimental \cite{Madelung} values 3.25 \AA, 5.20 \AA, and 0.382).

Defects were modeled using the supercell methodology. \cite{VanDeWalle_JAP_2004}
The defect calculations used 96-atom wurtzite supercells, with the lattice parameters
optimized at the HSE level. In the calculation of formation energies of charged defects,
as well as charge-state transition levels (ionization potentials), finite-size corrections as proposed
in Ref.~ \onlinecite{Freysoldt_PRL_2009} were included. The Brillouin zone was sampled at one special
$k$-point.\cite{Baldereschi_PRB_1973} For test systems, these calculations produce results for defect
levels within 0.03 eV of those obtained with a $2$$\times$$2$$\times2$ mesh.

While most of our calculations were performed using the PAW methodology, it makes calculations of
electron-phonon coupling matrix elements quite cumbersome. Such calculations are greatly facilitated
within the plane-wave pseudopotential (PW-PP) formalism \cite{Ihm_JPC_1979}, which we adopted for
this purpose. Norm-conserving Troullier-Martins pseudopotentials (PPs) \cite{Troullier_PRB_1991}
were generated at the PBE level using the {\sc fhi98pp} program. \cite{Fuchs_CPC_1999}
$3d$ states were included in the valence for both Zn and Ga. The energy cutoff for plane-wave expansion
of wavefunctions was set to 80 Ry in GaN and 100 Ry in ZnO. We used the {\sc cpmd} code, \cite{CPMD}
with the implementation of hybrid functionals discussed in Refs.~ \onlinecite{Todorova_JPCB_2006}, 
~\onlinecite{Broqvist_PRB_2009}, and ~\onlinecite{Komsa_PRB_2010}. Brillouin-zone sampling in these calculations was performed
using a single $\Gamma$ point. In order to reproduce experimental band gaps, $\alpha$ values of 0.38
for GaN and 0.47 for ZnO had to be used in these PP calculations, i.e., larger than in the PAW
calculations. We attribute this to the generation of PPs at the PBE level, rather than consistently
with hybrid functionals (cf. Refs.~ \onlinecite{Alkauskas_PRL_2008,Broqvist_PRB_2009,Komsa_PRB_2010,Alkauskas_PRL_2008b,
Wu_PRB_2009}). However, for parameters for which direct comparisons can be made, such as total energy
differences, equilibrium atomic configurations, or vibrational frequencies, the PW-PP calculations are in
gratifyingly good agreement with the PAW results; for instance, charge-state
transition levels for the defects considered here differ by 0.09 eV or less.


\subsection{Derivation of the capture coefficient \label{der}}

Let us consider a hole capture process at a single acceptor, as in Fig.~\ref{SRH}; the discussion
can be easily adapted to other cases. Let $V$ be a large volume that contains $P$ holes, their density
being $p=P/V$, and $M_A^{-}$ the total number of hole-capturing defects in the appropriate negative
charge state, with a density of $N_A^{-}$=$M_A^{-}/V$. The total density of defects is $N_A=N_A^{0}+N_A^{-}$.
Under non-equilibrium steady-state conditions, both electrons and holes can be present in the system.
Mobile carriers screen the Coulomb potential of impurities, with a screening length $\lambda$.
(For neutral impurities, $\lambda$ would be the extent of their short-range potential.) A few
distances $\lambda$ away from each impurity the potential essentially vanishes. We will assume
that $\lambda^3 N_A \ll 1$, implying that the region where the potential is \emph{not} negligible
constitutes a very small part of the solid. Since the hole density near the impurity is obviously
different from $p$, this assumption means that the hole density in the space where the potential
of impurity atoms can be neglected is equal to the average density, i.e., $p$. A similar condition
$\lambda^3 p \ll 1$ (i.e., $\lambda \ll p^{-1/3}$) allows us to assume that two holes do not
interact with the same impurity at the same time.

Computationally the most convenient quantity to calculate is the capture rate of one hole at one defect in
the whole volume $V$. Let the capture rate for such process be $r$ ($[r]$=s$^{-1}$). The capture rate of $P$
holes at all identical $M_A^{-}$ defects (all in their negative charge states) in volume $V$ is then
$\gamma_p$=$r M_A^{-} P$ ($[\gamma]$=s$^{-1}$). We can rewrite this equation as
$(\gamma_p/V)=rV\times(M_A^{-}/V)\times(P/V)=(rV)N_A^{-}p$. By comparing this equation with Eq.\
(\ref{rate}), and noting that, by definition $\gamma_p/V=R_p$ is the capture rate per unit volume,
we see that the hole capture coefficient is given by:
\begin{equation}
C_{p} = Vr.
\label{F1}
\end{equation}
From now on consider only one hole being captured by one defect.

The general idea behind nonradiative processes due to multiphonon emission is closely
related to the concept of electron-phonon coupling in bulk solids. \cite{Hedin_1970}
We briefly review the main ideas, emphasizing the aspects specific to defects.
The many-body Hamiltonian of the entire system of electrons and ions is
\begin{equation}
\hat{H}=\hat{T}_I + \hat{T}_e + \hat{V}_{II} + \hat{V}_{ee}  + \hat{V}_{Ie},
\label{Schrodinger}
\end{equation}
where $\hat{T}$ represents kinetic energy, $\hat{V}$ represents Coulomb interaction, and the subscript ``\emph{e}''
is for electrons and ``\emph{I}'' for ions. For an isolated system at zero temperature the solution of
the Schr\"{o}dinger equation $\hat{H}\Psi_n=E_n\Psi_n$ yields the energy spectrum $E_n$ and many-body wavefunctions
$\Phi_n\left(\{Q\},\{x\}\right)$. $\{x\}$ represents all electronic degrees of freedom and $\{Q\}$ represent all
ionic coordinates (which can be transformed to phonon coordinates in the harmonic approximation). In most
practical situations, however, it is more useful to describe the system not via eigenstates of the full
Hamiltonian $\hat{H}$, but via eigenstates of a simpler Hamiltonian $\hat{H}_0$ that encodes the essential physics of
the system.\cite{Hedin_1970} Eigenstates of $\hat{H}_0$ can be written as a product of the electronic
and the ionic part. The term $\Delta \hat{H}=\hat{H}-\hat{H}_0$ is then the perturbation that causes transitions between
eigenstates of $\hat{H}_0$. These transitions should be rare in order to ensure that the Hamiltonian $\hat{H}_0$
captures the essential physics of the system.\cite{Hedin_1970} The part of $\Delta\hat{H}$ that is due to
the ions and that induces transitions between different electronic states, such as in the case of nonradiative
carrier capture, is the electron-phonon coupling $\Delta \hat{H}^{e-ph}$. The remaining piece describes
electron-electron and phonon-phonon interactions that are not discussed further.

The time scale associated with carrier capture processes in semiconductors is usually much larger than both
phonon lifetimes and periods of lattice vibrations (an assumption that has to be verified \emph{a posteriori}).
As a result, the most convenient starting point to describe a coupled system of electrons and ions is the so-called
static approximation. \cite{Paessler_pssb_1975,Peuker_pssb_1982,Goguenheim_JAP_1990} In this approximation, which we will adopt here,
the total wavefunction of the system can be written as $\Psi \left(\{Q_0\},\{x\}\right) \chi(\{Q\})$,
where $\Psi \left(\{Q_0\},\{x\}\right)$ is the electronic wavefunction calculated for a chosen ionic configuration
$\{Q_0\}$, and $\chi(\{Q\})$ is the ionic wavefunction. The choice of $\{Q_0\}$ will be discussed in Sec.\ \ref{initial}.

Let the many-body electronic wavefunction that describes a hole in the valence band (which is perturbed by the
presence of the defect) and a negatively charged defect be $\Psi_i \left(\{Q_0\},\{x\}\right)$. This is the
excited (initial) electronic state. The electronic wavefunction that describes a hole trapped on a defect (yielding
a neutral charge state of the center) is $\Psi_f \left(\{Q_0\},\{x\}\right)$. This is the ground (final)
electronic state. The associated ionic wavefunctions are $\chi_{im}\left(\{Q\}\right)$ and $\chi_{fn}\left(\{Q\}\right)$,
where $n$ and $m$ are quantum numbers for ionic states.

At finite temperatures $T$ free holes occupy various electronic states according to the Fermi-Dirac or, in
the non-degenerate case, the Boltzmann distribution. As a result, they cannot be described by a single initial
state $\Psi_i \left(\{x\}\right)$. The carrier capture rate that is experimentally measured is the weighted average
over all initial electronic states. We adopt an approximation that charge carriers can be represented by a single
initial electronic state; see Chapter 14.3 of Ref.~\ \onlinecite{Stoneham} for a more in-depth discussion. In the
case of the non-degenerate hole or electron gas, this special state represents particles with a thermal velocity;
in the degenerate case the special state represents particles at the Fermi surface.

In this work, we consider the interaction within the first order of electron-phonon coupling.
Under this assumption, the capture rate $r$ that enters into Eq.\ (\ref{F1}) is given by Fermi's golden rule
(see, e.g., Sec. 14.2 of Ref.~ \onlinecite{Stoneham}):
\begin{equation}
r = \frac{2\pi}{\hbar} g \sum_{m} w_{m} \sum_{n} \left|\Delta H^{e-ph}_{im;fn}\right|^2 \delta (E_{im}-E_{fn}).
\label{F2}
\end{equation}
Here $w_{m}$ is the thermal occupation of the vibrational state $m$ of the excited electronic state, and
$E_{im}$ and $E_{fn}$ are total energies of the initial and the final vibronic state. $g$ is the degeneracy
factor of the final state; it reflects the fact that there might exist a few equivalent energy-degenerate
(or nearly degenerate) atomic configurations of the final state. For example, the neutral charge
state of Li$_{\text{Zn}}$ in ZnO can correspond to four different lattice relaxations in which the hole
is localized on one of the four surrounding oxygens,\cite{Du_PRB_2009,Carvalho_PRB_2009,Bjorheim_JPCC_2012,Lyons_2014}
yielding $g=4$. Similarly, $g=4$ for GaN:C$_{\text{N}}$.\cite{Lyons_APL_2009} We do not distinguish between
axial and azimuthal configurations in the wurtzite structure. In the equation above, $\Delta H^{e-ph}_{im;fn}$
is the electron-phonon coupling matrix element. In the static approach,\cite{Paessler_CJP_1982,Paessler_pssb_1975}
$\Delta \hat{H}^{e-ph}=\hat{H}\left(\{Q\},\{x\}\right)-\hat{H}\left(\{Q_{0}\},\{x\}\right)$.

To make the problem more tractable, subsequent approximations need to be employed. The first of
those is the linear-coupling approximation. \cite{Stoneham} In this approximation $\Delta \hat{H}^{e-ph}$
is Taylor-expanded in $\{Q\}$ around $\{Q_{0}\}$ (see Sec.\ \ref{initial}), and only the first-order
terms are retained. The matrix element $\Delta H_{im;fn}^{e-ph}$ is then given by:
\begin{equation}
\Delta H_{im;fn}^{e-ph} = \sum_{k}
\underbrace{\left \langle \Psi_i \left | \partial \hat{H} /\partial Q_{k} \right| \Psi_f \right \rangle}_{W^k_{if}}
\langle \chi_{im} \left | Q_k-Q_{0; k}\right | \chi_{fn} \rangle
\label{F3}
\end{equation}
The sum runs over all phonon modes $Q_k$, and $Q_{0; k}$ is the projection of the initial atomic configuration
$\{Q_0\}$ along each of the phonon coordinates. $W^k_{if}$ is the electron-phonon coupling matrix element
pertaining to the phonon $k$. Equations\ (\ref{F1}), (\ref{F2}) and (\ref{F3}) form the starting point
for our computational determination of $C_p$.

\subsection{Vibrational problem \label{vib-pr}}

The approximations introduced so far are fairly standard and have been employed in
previous work. \cite{Abakumov,Paessler_pssb_1975,Peuker_pssb_1982,Goguenheim_JAP_1990}
Here we introduce an additional approximation, relating to the phonon coordinates,
which will turn out to be essential for making the calculations of electron-phonon matrix elements
feasible. In particular, we will consider only {\it one} special phonon mode that replaces the sum over all
vibrational degrees of freedom in Eq.\ (\ref{F3}). The choice of the phonon mode is motivated by the following
reasoning. We are dealing with deep levels, with ionization energies $\Delta E$ that are usually many times
larger than the energy of the longitudinal optical phonon (which has the largest energy of all phonon modes). This
is the reason why a single phonon process is not sufficient to couple the two electronic states, and
a MPE is necessary. The phonons that contribute most to the the sum in Eq.\ (\ref{F3}) are those that
couple most strongly to the distortion of the defect geometry during the carrier capture process.
This is ensured by the second factor in Eq.\ (\ref{F3}), since this expression vanishes for those
modes that do not couple to this distortion.

This approach is supported by results that have been obtained in the case of radiative transitions.
In the case of luminescence at defects with strong electron-phonon interactions, as quantified by
their Huang-Rhys factors \cite{Huang_PRCL_1950} $S\gg1$, it is possible to show numerically that
replacing many participating phonon coordinates with one carefully chosen effective phonon mode is an excellent
approximation. \cite{Alkauskas_PRL_2012} This conclusion is in line with an empirical finding that it is often
possible to describe the temperature dependence of broad defect luminescence bands considering only one
effective vibrational degree of freedom \cite{Stoneham,Reshchikov_JAP_2005}. This special mode corresponds to
an effective vibration \cite{Schanovsky_JVST_2011,Alkauskas_PRL_2012} where the displacement of an atom $\alpha$
along the direction $t$=$\{x,y,z\}$ is proportional to $\Delta R_{\alpha t}$=$R_{i; \alpha t} - R_{f; \alpha t}$,
where $R_{\{i,f\}; \alpha t}$ are atomic coordinates in the equilibrium configuration of the excited (initial)
and the ground (final) state. In this one-dimensional model the generalized configuration coordinate
$Q$ for values of atomic positions $R_{\alpha t}$ that correspond to this displacement is
\begin{equation}
Q^2 = \sum_{\alpha, t} m_{\alpha}\left(R_{\alpha t} - R_{f; \alpha t}\right)^2,
\label{vib1}
\end{equation}
where $m_{\alpha}$ are atomic masses. The geometry of the ground state (final state $f$) corresponds to $Q$=0, while the geometry
of the excited state (initial state $i$) corresponds to $Q$=$\Delta Q$ with
\begin{equation}
\left(\Delta Q\right)^2 = \sum_{\alpha, t} m_{\alpha} \Delta R_{\alpha t}^2,
\label{vib2}
\end{equation}
In this description the configuration coordinate of Eq.\ (\ref{vib1}) has units of amu$^{1/2}$\AA $ $
(amu - atomic mass unit). We will give a brief description of changes of the defect geometry
encoded in $\Delta Q$ when discussing specific systems in Section \ref{res}.
The plot that shows the total energies in the ground and the excited states
$E_{\{i,f\}}$ as a function of $Q$ is called the configuration coordinate diagram (cc diagram) \cite{Stoneham};
we have shown a schematic example in Fig.~\ref{SRH}(b). The frequency of the effective vibration in the
ground and the excited state is given as
\begin{equation}
\Omega^2_{\{i,f\}} = \frac{\partial ^2 E_{\{i,f\}}}{\partial Q^2}.
\label{vib3}
\end{equation}
An auxiliary quantity $\left(\Delta R\right)^2$=$\sum_{\alpha, t}\Delta R_{\alpha t}^2$,
allows to define the modal mass of the vibration via $\Delta Q$=$M^{1/2}\Delta R$. \cite{Schanovsky_JVST_2011}
The knowledge of $M$ is useful for interpreting the value of $\Omega$ for different defects.\cite{Alkauskas_PRL_2012}
A very useful dimensionless quantity is the Huang-Rhys factor, defined as \cite{Huang_PRCL_1950,Alkauskas_PRL_2012}
\begin{equation}
S_{\{i,f\}}=\frac{1}{2\hbar}(\Delta Q)^2 \Omega_{\{i,f\}}.
\label{HR4}
\end{equation}
The case $S\gg1$ corresponds to large lattice relaxations associated with the change of the charge state.
We note that the special mode $Q$ is not an eigenstate of the vibrational Hamiltonian, but it serves as a
very useful approximation and has a clear physical meaning. Possible errors introduced by the use of the
one-dimensional approximation are critically reviewed in Sec.\ \ref{multi-1D}.

\subsection{Electron-phonon matrix elements \label{e-ph-matrix}}

Thanks to the one-dimensional (1D) approximation described in Sec.~\ref{vib-pr} we have to determine only
a single electron-phonon coupling matrix element:
\begin{equation}
W_{if}=\Braket {\Psi_{i} |\partial \hat{H}/\partial Q | \Psi_{f}}.
\label{e-ph1}
\end{equation}
At this stage, $\Psi_{\{i,f\}}$ are still many-electron wavefunctions, and $\hat{H}$ is the many-body
Hamiltonian of the system. In an independent-particle picture corresponding to the (generalized)
Kohn-Sham approach of DFT, we will assume that the many-body Hamiltonian and many-electron wavefunctions in
Eq.\ (\ref{e-ph1}) can be replaced by their single-particle counterparts $\hat{h}$ and $\psi_{\{i,f\}}$, i.e.:
\begin{equation}
W_{if}=\Braket {\psi_{i} |\partial \hat{h}/\partial Q | \psi_{f}}.
\label{e-ph2}
\end{equation}
Whereas wavefunctions $\Psi_{\{i,f\}}$ describe the entire electronic system, single-particle wavefunctions
$\psi_{\{i,f\}}$ have a different meaning: $\psi_i$ corresponds to the hole in the valence band perturbed
by the presence of the defect, and $\psi_f$ is the localized defect state. Indeed, for perturbation
theory to be physically meaningful, both states $\psi_{\{i,f\}}$ have to be eigenstates of the same
Hamiltonian: the initial state has to correspond to the perturbed hole state rather than a hole state
in an unperturbed bulk material.

To calculate electron-phonon matrix elements we use hybrid functionals within the PW-PP approach,
as discussed in Sec.~\ \ref{Comp}. Therefore, $\hat{h}$ contains nonlocal Fock exchange, as well as
the nonlocal part of pseudopotentials. These terms would have to be calculated explicitly if
Eq.\ (\ref{e-ph2}) were used. To avoid such a cumbersome procedure, it is extremely convenient
to use an alternative expression that follows directly from perturbation theory
[e.g., Eq.\ (28) in Ref.~\ \onlinecite{Baroni_RMP_2001}]:
\begin{equation}
W_{if}=
\Braket {\psi_{i} | \frac{\partial \hat{h}}{\partial Q} | \psi_{f}}=
\left(\varepsilon_{i}-\varepsilon_{f} \right)
\Braket { \psi_{i} | \frac{\partial \psi_{f}} {\partial Q} }
\label{e-ph3}
\end{equation}
In this expression the main effort in calculating the matrix element boils down to the calculation of the
derivative $\partial \psi_{f}/\partial Q$.  This is accomplished by evaluating the derivative numerically
via finite differences, as discussed for specific defects in Sec.~\ref{initial}.

\subsection{Bulk scattering states \label{scatter}}

The methodology outlined in Sec.\ \ref{der} above relies on calculating the capture rate $r$ for one hole,
with a finite velocity, at one defect in the entire (large) volume $V$. The role of all other carriers is to
screen the long-range Coulomb interaction between the hole and the defect. The electron-phonon coupling matrix
element for one special phonon mode is determined via Eq.~(\ref{e-ph3}), and the capture coefficient $C_p$ is
subsequently determined via Eqs.\ (\ref{F1}), (\ref{F2}) and (\ref{F3}). Only one phonon coordinate is
retained in expression (\ref{F3}).

The problem with this formulation is the following. Actual calculations are performed for a
system with a relatively small volume, the supercell with a volume $\tilde{V}$ that is constrained
by computational limitations. While there is plenty of evidence that the localized defect state $\psi_f$ is
accurately represented in such supercell calculations, this is not necessarily the case for the initial perturbed
bulk state $\psi_i$. In particular, if the capturing center is charged, the screened Coulomb interaction between
the defect and the carrier significantly affects the capture processes. Such interactions are not well represented
in the supercell calculation. Let us picture, as an example, a charged carrier with a vanishing kinetic energy
being captured at a repulsive center. As the size of the system $V$ grows, the particle is 
expelled further and further away from the defect. In the limit of an infinite volume $V$, and zero kinetic energy
of the charge carrier, the capture rate would tend to zero. However, in the supercell of volume  $\tilde{V}$
the carrier cannot be expelled to infinity, and the capture rate remains finite, which is an incorrect
physical result. Similar considerations also apply to attractive centers and emphasize the need of
the correction term, which we discuss here.

Let the electron-phonon coupling matrix element calculated in the computational supercell be $\tilde{W}_{if}$.
It is calculated via the equation, similar to Eq.~(\ref{e-ph3}):
\begin{equation}
\tilde{W}_{if}=
\Braket {\tilde{\psi}_{i} | \frac{\partial \hat{h}}{\partial Q} | \psi_{f}}=
\left(\varepsilon_{i}-\varepsilon_{f} \right)
\Braket { \tilde{\psi}_{i} | \frac{\partial \psi_{f}} {\partial Q} }
\label{e-ph8}
\end{equation}
Here, $\tilde{\psi}_i$ is the bulk wavefunction in the supercell of volume $\tilde{V}$,
chosen to be at the $\Gamma$ point of the supercell. Let the corresponding carrier capture coefficient, determined via
equations analogous to Eqs.\ (\ref{F1})-(\ref{F3}), whereby all the parameters of the real system are substituted with
corresponding values from the supercell calculation, be $\tilde{C}_{p}$. Similar to the procedure proposed in
Refs.~\ \onlinecite{Bonch-Bruevich_FTT_1959} and~ \onlinecite{Paessler_pssb_1976} we express the actual capture
coefficient as
\begin{equation}
C_{p} = f\left(n,p,T\right)\tilde{C}_{p},
\label{sca1}
\end{equation}
where $f\left(n,p,T\right)$ is a dimensionless scaling factor that depends on the reference calculation
used to determine the matrix element $\tilde{W}_{if}$, the charge state of the defect, as well as environmental
parameters: electron density $n$, hole density $p$, and temperature $T$. Bonch-Bruevich \cite{Bonch-Bruevich_FTT_1959}
and later P\"{a}ssler \cite{Paessler_pssb_1976} provided analytic expressions of $f(n,p,T)$ for both repulsive
and attractive centers. In the present Section we derive an expression of $f(n,p,T)$ in the context of our supercell
approach. Our analysis follows that of P\"{a}ssler,\cite{Paessler_pssb_1976} but is adapted for use in conjunction
with supercell calculations of defects.

The function $f(n,p,T)$ can in principle be constructed using a first-principles approach. However, such a
calculation would be very cumbersome and not particularly useful at this point, keeping in mind that other,
more limiting approximations have already been made. Instead we employ a model calculation to determine $f(n,p,T)$.

Let us assume that the perturbed bulk wavefunction $\psi_i$ in the real physical system can be described as a product
of the wavefunction that reflects the atomic-scale behavior $\xi_i$ and the envelope wavefunction $\phi_i$ that changes
on a scale larger than the unit cell: $\psi_i=\xi_i \phi_i$. We chose the normalization condition for $\phi_i$ to be the
same as for $\psi_i$. $\xi_i$ is a fast-varying dimensionless function.
Such a description is in the spirit of the effective-mass
approximation.\cite{Kohn_PR_1957} The electron-phonon coupling matrix element $W_{if}$ can then be
expressed as
\begin{equation}
W_{if}\approx
\phi_i(0)\Braket { \xi_{i} | \frac{\partial \hat{h}}{\partial Q} |\psi_{f}}=\phi_i(0)w_{if},
\label{sca19}
\end{equation}
where $\phi_i(0)$ is the value of the envelope wavefunction at the defect site, and a new matrix
element $w_{if}$  was introduced. According to the methodology described in Secs.\ \ref{der},
\ref{vib-pr}, and \ref{e-ph-matrix} [Eqs.\ (\ref{F1}), (\ref{F2}), (\ref{F3}), (\ref{e-ph2}), and (\ref{sca19})]
the capture coefficient is then proportional to
\begin{equation}
C_p \sim  V \left | \phi_i(0) \right |^2 |w_{if}|^2.
\label{sca2}
\end{equation}
Here, $V$ is the large volume of the material introduced in Sec.~\ref{der}.
In the region where the potential of impurities is negligible $|\phi_i|=1/\sqrt{V}$.
Because of the interaction with the impurity, $|\phi_i(0)|$ can have a different value.

Let us assume that the perturbed bulk state in the computational supercell can also be written in terms
of a similar product, i.e., $\tilde{\psi}_i=\tilde{\phi}_i\xi_i$. Because of its localized nature, the
defect wavefunction $\psi_f$ is the same in the supercell of volume $\tilde{V}$ as in a large volume
$V$. By definition, the same holds for the ``atomic'' part of the bulk wavefunction  $\xi_i$.
As a result,
\begin{subequations}
\begin{align}
\tilde{W}_{if} = \Braket { \tilde{\psi}_{i} | \frac{\partial \hat{h}}{\partial Q} |\psi_{f}} \label{e-ph-tilde} \\
\approx\tilde{\phi}_i(0)\Braket {\xi_{i}|\frac{\partial \hat{h}}{\partial Q} |\psi_{f} } &= \tilde{\phi}_i(0)w_{if}\label{e-ph-tilde1}
\end{align}
\end{subequations}
Accordingly:
\begin{equation}
\tilde{C}_p \sim\tilde{V} |\tilde{\phi}_i(0)|^2 |w_{if}|^2.
\label{sca3}
\end{equation}
Therefore, from Eqs.\ (\ref{sca1}), (\ref{sca2}), and (\ref{sca3}):
\begin{equation}
f\left(n, p, T\right)= \frac{V \left|\phi_i(0)\right|^2} {\tilde{V}|\tilde{\phi}_i(0)|^2}.
\label{sca4}
\end{equation}

In practice, we use the following procedure. The value $V[\phi_i(0)]^2$ is determined
by considering a scattering problem for a particle with a finite momentum $k$, \cite{Landau}
which we take to be the thermal momentum $k_T$ for the non-degenerate case. Far from the scattering
center the wavefunction is normalized as required by the formulation of our problem
($V[\phi_i(\mathbf{r})]^2=1$ for $r\rightarrow\infty$). Within the $s$-wave approximation,\cite{Landau}
the value of the wavefunction at the origin is determined by a numerical integration of the
Schr\"{o}dinger equation for the $l$=0 angular momentum component of the scattering wavefunction
with an asymptotic form that corresponds to our normalization. The scattering potential that we use is
\begin{equation}
\label{eq:V}
V(r)=\frac{Z}{\varepsilon_0 r}\text{erf}(r/r_0) \exp(-r/\lambda).
\end{equation}
Here $\varepsilon_0$ is the low-frequency dielectric constant of the host material, $r_0$ is the extent of
the defect wavefunction, and $\lambda$ is the screening length due to the presence of other charge carriers,
as discussed at the beginning of Sec.\ \ref{der}. In the case of a non-degenerate gas we use the Debye-H\"{u}ckel
screening length that depends on $T$ and on the carrier densities $n$, $p$, explaining the overall dependence
of $f$ on these parameters. At room temperature, the hole gas is non-degenerate up to densities of
$p \sim10^{19}$cm$^{-3}$ in both GaN and ZnO.

The value of $V[\phi_i(0)]^2$ is obtained by consideration of the scattering problem with the potential in
Eq.~(\ref{eq:V}). We determine $r_0$ by comparing the behavior of the bulk wavefunction in the presence of
a charged defect in the actual supercell calculation with the wavefunction obtained from a model supercell
calculation within the effective-mass approximation. $r_0$ is chosen so that the behavior of the envelope
wavefunction in the model supercell calculation accurately represents the behavior of the envelope function
in a real calculation. This model supercell calculation also yields the value of $\tilde{V}|\tilde{\phi}_i(0)|^2$.

For an attractive Coulomb potential the problem can be solved analytically.\cite{Landau,Paessler_pssb_1976}
When $k\ll1/a_{B}^{*}$, where $a_{B}^{*}$ is the effective Bohr radius in the material,
$f(k)\sim1/k$ (see Eq.~(4.4) in Ref. ~\onlinecite{Paessler_pssb_1976}). We find numerically
that a similar form  $f=A/k$ describes very accurately the behavior of carriers also for the potential 
of the form Eq.~(\ref{eq:V}) when the center is attractive ($Z<0$). We find that $f$
depends very weakly on $\lambda$, which in its turn depends on carrier density, and thus we can use
density-independent scaling function.
In our formulation $k$ is the average thermal momentum of holes and $A$ is a constant.
For the non-degenerate hole gas $\hbar k=(3k_{B}mT)^{1/2}$ ($k_B$ is the Boltzmann constant), and thus the
scaling function depends only on temperature:
\begin{equation}
f(T)=\frac{C}{T^{1/2}},
\label{temp-scaling}
\end{equation}
where $C$ is a constant determined numerically. For the two attractive centers considered in the
present study (GaN:C$_{\text{N}}$ and ZnO:Li$_{\text{Zn}}$) we found $C\approx150$ K$^{1/2}$
when the electron-phonon matrix element $\tilde{W}_{if}$ is determined for a neutral charge state
for reasons discussed in Sec.\ \ref{initial}. In the scattering problem,
we assumed effective hole masses $m_h$=$1.0$ for GaN\cite{Pankove_1975,Santic_SST_2003}, and
0.6 for ZnO.\cite{Hummer_PSS_1973} 
For attractive centers and temperatures considered in this work ($T<1000$ K) $f>1$.

This result is intuitive and can be explained as follows.  Close to the defect the wavefunction of the hole
has a larger amplitude with respect to its asymptotic value far away from the defect; in the classical
reasoning, the hole spends more time near the defect due to Coulomb attraction. The function $f$ reflects this
enhancement. For example, the factor $f$ is about 10 at room temperature.

The third defect considered in our work, GaN:(Zn$_{\text{Ga}}$-$V_{\text{N}}$), captures holes
in a neutral charge state (see Sec.\ \ref{ZnGa-VN}), thus there are no long-range
Coulomb interactions between the defect and the hole. However, the electron-phonon coupling matrix
element is calculated in the positively charged state (96-atom supercell),
as discussed in Sec.\ \ref{initial}. In this case we find $f=1.05$. This implies that in the supercell
calculation the hole is repelled from the defect more than in the actual situation.

In the case of repulsive centers $f$ depends sensitively both on temperature ($f\sim\exp(-a/T^{1/3}$)
\cite{Bonch-Bruevich_FTT_1959}, and on the density of charge carriers. Repulsive centers are not considered
in this work.

\subsection{Initial state for perturbation theory \label{initial}}

The actual quantity that is calculated is the capture coefficient $\tilde{C}_p$ that
corresponds to our computational setup. The expression can be derived from
Eqs.~(\ref{F1}), (\ref{F2}), and (\ref{F3}), whereby all quantities correspond
to the parameters in the supercell calculation (rather than the actual system)
and only one phonon mode is retained in Eq.~(\ref{F3}):
\begin{eqnarray}
\tilde{C}_p=\tilde{V}\frac{2\pi}{\hbar}g\tilde{W}_{if}^2\sum_m w_m \sum_n \left| \left \langle
\chi_{im}|Q-Q_{0}|\chi_{fn}\right \rangle \right |^2
\nonumber \\
\times\delta (\Delta E + m\hbar\Omega_{i}-n\hbar\Omega_{f}).
\label{r5}
\end{eqnarray}
$\tilde{W}_{if}$ is given via Eq.~(\ref{e-ph8}).
For numerical evaluation, the $\delta$ function is replaced by a smearing function of finite width, a practice
also employed in calculating luminescence lineshapes. \cite{Alkauskas_PRL_2012} In this section we address the
following questions: (i) which atomic configuration $\{Q_0\}$ should we choose as a starting point for perturbation
theory and (ii) for which charge state should we calculate the electron-phonon matrix element $\tilde{W}_{if}$
in Eqs.\ (\ref{e-ph8}) and (\ref{r5})?

During a nonradiative process the carrier in a delocalized state is captured to a localized defect
state. Thus, in the configuration $\{Q_0\}$ a single-particle defect level should be well defined
and be in the bulk band gap. This is the single most important criterion for the choice of $\{Q_0\}$.
Let us consider acceptor defects GaN:C$_{\text{N}}$ and ZnO:Li$_{\text{Zn}}$ as an example.
As before, we study the capture of a hole by a negatively charged acceptor [process (1) in Fig.\
\ref{SRH}(b)]. Actual first-principles calculations show that in the case of neutral acceptors in their
equilibrium geometries there is indeed one clearly distinguishable empty defect level in the band gap,
representing a trapped hole. In contrast, in the case of the negatively charged defect in its equilibrium
geometry the supercell calculation produces one or more diffuse single-particle defect states that have
moved down in energy and that couple strongly with bulk states. However, when the calculation is performed
for the same negatively charged acceptor but rather in the equilibrium geometry of the {\it neutral} system,
one doubly-occupied single-particle defect state moves up in energy and into the band gap.
A defect state to which the hole is being captured can be clearly identified again. The bottom line is
that, when $\{Q_0\}$ corresponds to the equilibrium geometry of the neutral
charge state, a single-particle defect state can be clearly identified in both the neutral and the
negative charge states. This choice of $\{Q_0\}$ thus  yields good single-particle wavefunctions
for perturbation theory.

For this particular choice of $\{Q_0\}$ the electron-phonon coupling constant $\tilde{W}_{if}$ can be then
calculated for either the neutral or the negatively charged state. We find that the $\tilde{W}_{if}$ values
calculated for the two states differ by about 5\%. However, a different charge state for the
calculation of the electron-phonon coupling matrix element yields a different scaling function $f$,
as discussed in Sec.\ \ref{scatter} [Eq.\ (\ref{sca4})]. In the end, the calculated
capture rates are within 1\% of each other. This result is reassuring, and also tells us something about the physics:
the defect wavefunction does not change much when the defect state is filled with an electron.

If the band structure is such that the highest occupied states correspond to several closely spaced valence
bands (which is the case for the most commonly used semiconductors) attention needs to be devoted to the choice
of the valence band that represents the hole wave function. For example, the highest occupied states at the
zone center of wurtzite-phase semiconductors, such as GaN and ZnO, consist of the heavy-hole (HH), light-hole (LH),
and crystal-field split-off (CH) bands.\cite{Piprek} The splitting between LH and HH is mainly due to the spin-orbit
interactions, and is only a few meV for these two materials. Crystal-field effects are larger, and the CH band is
$\sim$20 meV below the valence-band maximum (VBM) in GaN, and $\sim$60 meV in ZnO. \cite{Yan_SST_2011}
Strain or effects of confinement in quantum wells could modify the splitting and ordering of these bands,
and in an actual sample, the density of holes in each band is determined by the thermal occupation.
For example, in bulk ZnO the CH band will be much less populated with holes than the other two bands at room
temperature, and this can be relevant experimentally (see Sec.\ \ref{Li}). Since we do not know {\it a priori}
which valence band(s) will play the most important role in specific experimental situations, we explicitly
calculate the electron-phonon coupling to all three valence bands in the supercell.

The use of the supercell itself introduces an additional complication, since the splitting between the bands and
their ordering can be significantly affected by the defect. We find that the valence band that interacts most
strongly with the defect state is always pushed below the other two bands.
However, while the precise energetic position of the bands may be affected, we find that the {\it character}
of the valence bands is generally retained in defect supercells, allowing us to meaningfully calculate the
electron-phonon coupling matrix elements for the separate valence bands. While these matrix elements could in
principle be explicitly employed in calculations that reflect specific experimental conditions,
for purposes of reporting our results in the present paper the matrix element of Eq.\ (\ref{e-ph1})
that enters into the final calculations is defined as the mean-square average of the three separate matrix
elements.

\begin{figure}
\includegraphics[width=8.5cm]{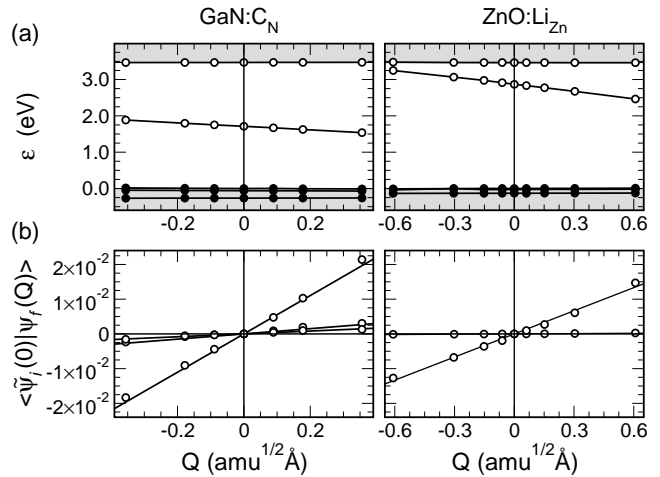}
\caption{Calculation of the electron-phonon matrix element $\tilde{W}_{if}$ using Eq.\ (\ref{e-ph8})
for GaN:C$_{\text{N}}$ and ZnO:Li$_{\text{Zn}}$.
(a) Eigenvalues of defect and bulk wavefunctions as a function of $Q$.
(b) The overlap $\Braket { \tilde{\psi}_{i}(0)| \psi_{f}(Q)}$ as a function of $Q$.
}
\label{e-ph4}
\end{figure}

Here, we illustrate the calculation of $\tilde{W}_{if}$ for two specific defects, C$_\text{N}$ in GaN
and Li$_{\text{Zn}}$ in ZnO. The calculation of the electron-phonon matrix element $\tilde{W}_{if}$ using
Eq.\ (\ref{e-ph8}) is shown in Fig.\ \ref{e-ph4}. Panel (a) shows the single-particle eigenvalues
as a function of a generalized coordinate $Q$, where $Q_0$, corresponding to the geometry of the neutral
defect state, is set to 0. The eigenvalues are referenced to the VBM. In the case of a defect immersed in
infinite bulk the eigenvalues of bulk states should not be dependent on $Q$. This condition is fulfilled in
our supercell calculations [Fig.\ \ref{e-ph4}(a)]. In contrast, the defect state shows a pronounced linear
dependence on $Q$. For the calculation of $\tilde{W}_{if}$ the value of $\varepsilon_f-\varepsilon_i$ at $Q=0$
is taken.

In Fig.\ \ref{e-ph4}(b) the overlap integral $\Braket {\tilde{\psi}_{i}(0)|\psi_{f}(Q)}$ is plotted as a
function of $Q$ for all three valence bands. The derivative $\Braket { \tilde{\psi}_{i} | \partial \psi_{f}/\partial Q}$
used for the calculation of the matrix element in Eq.~(\ref{e-ph8}) was determined from a linear fit to this dependence.
The coupling to one of the three valence bands can be as much as two orders of magnitude larger than for the
other bands, as discussed above.

\subsection{Brief summary of the methodology \label{briefs}}

To recap, we determine the carrier capture rate that is specific to our supercell geometry using
Eq.\ (\ref{r5}). $\tilde{V}$ is the volume of the supercell; $g$ is the degeneracy of the final state;
$\tilde{W}_{if}$ are electron-phonon coupling matrix elements, given in Eq.~(\ref{e-ph8}) [cf. Fig.~\ref{e-ph4}];
$\Delta E$ is the energy difference between the ground and excited state that is given by the position
of the charge-state transition level above the VBM\cite{VanDeWalle_JAP_2004}; $Q$=$Q_0$=$0$
is chosen to correspond to the equilibrium atomic configuration of the ground state;
the equilibrium atomic configuration of the excited state is offset by $Q$=$\Delta Q$ [Eq.~(\ref{vib2})].
All these quantities are summarized in Table~\ref{Table1}. In addition, we provide Huang-Rhys factors
$S_f$ [Eq.~(\ref{HR4})]. $\delta$-functions in the sum Eq.~(\ref{r5}) are replaced by Gaussians with widths
$\sigma$=$0.8\hbar\Omega_f$.\cite{Alkauskas_PRL_2012}

Finally, the actual carrier capture coefficient $C_p$ is obtained via $C_{p} = f\tilde{C}_{p}$
[Eq.~(\ref{sca1})] with the scaling function $f$. As discussed in Sec.~\ref{scatter}, the calculation of $f$ may
require a simulation in its own right; for the case of hole capture by a negatively charged defect, when the
reference system is that of the neutral charge state, we use the form Eq.~(\ref{temp-scaling}) for $f$. This is
the situation that occurs in the examples of GaN:C$_{\text{N}}$ and ZnO:Li$_{\text{Zn}}$, to be discussed in
Secs.~\ref{CN-res} and \ref{Li}. For the case of hole capture by a neutral defect, which
applies to GaN:(Zn$_{\text{Ga}}$-$V_{\text{N}}$) to be discussed in Sec.~\ref{ZnGa-VN},
the reference system is that of a positive charge state and $f=1.05$ for our particular supercell.

\begin{table*}
\caption{Key parameters for the three defects studied in this work: total mass-weighted distortions
$\Delta Q$ [Eq.\ (\ref{vib2})], ionization energies $\Delta E$,
energies of effective vibrations $\hbar\Omega_{\{i,f\}}$ [Eq.\ (\ref{vib3})] (charge state is given in parentheses),
Huang-Rhys factors for the final state [Eq.\ (\ref{HR4})], degeneracy factor $g$ of the final state [cf.\ Eq.\ (\ref{F2})),
electron-phonon coupling matrix elements $\tilde{W}_{if}$ [Eq.\ (\ref{e-ph8}) and Fig.\ \ref{e-ph4}; the
charge state of the defect for which the matrix element is calculated is shown in parentheses],
and volume of the supercell $\tilde{V}$ in first-principles calculations.}
\begin{ruledtabular}
\begin{tabular}{l c c c c c c c c c}
&$\Delta Q$ (amu$^{1/2}$\AA)&\multicolumn{2}{c}{$\hspace{7mm} $  $\Delta E$ (eV)}&$\hbar\Omega_{i}$ (meV)&$\hbar\Omega_{f}$ (meV)&$S_{f}$&$g$&$\tilde{W}_{if}$ (eV/amu$^{1/2}$\AA) & $\tilde{V}$ (\AA$^3$)\\
Defect                             &     &theory                   &expt.                &        &      & &  & &\\
\hline
GaN:C$_{\text{N}}$                 &1.61 &1.02 (this work)     &0.85\footnotemark[1] &42 ($-$) &36 (0) &10& 4& $6.4\times10^{-2}$ (0)&1100  \\

ZnO:Li$_{\text{Zn}}$               &3.22 &0.80\footnotemark[2], 0.49\footnotemark[3], 0.46\footnotemark[4]
&0.53\footnotemark[5] &36 ($-$)& 25 (0) &28& 4&$3.9\times10^{-2}$ (0)&1136 \\

GaN:(Zn$_{\text{Ga}}$-$V_{\text{N}}$)&3.33 &0.90\footnotemark[6], 0.88 (this work)   &$-$                  &26 (0)  &22 (+) &30&1& $1.0\times10^{-2}$ ($+$)&1100  \\

\end{tabular}
\label{Table1}
\end{ruledtabular}
\footnotetext[1]{Ref.\ \onlinecite{Demchenko_PRL_2013}}
\footnotetext[2]{Ref.\ \onlinecite{Carvalho_PRB_2009}}
\footnotetext[3]{Ref.\ \onlinecite{Bjorheim_JPCC_2012}}
\footnotetext[4]{Ref.\ \onlinecite{Lyons_2014}}
\footnotetext[5]{Ref.\ \onlinecite{Reshchikov_PB_2007}}
\footnotetext[6]{Ref.\ \onlinecite{Shi_PRL_2012}}
\end{table*}


\section{Results \label{res}}

To illustrate our methodology, we study two defects in GaN, namely carbon on the nitrogen site
(GaN:C$_{\text{N}}$) and a complex of zinc on a gallium site with a nitrogen vacancy
(GaN:(Zn$_{\text{Ga}}$-$V_{\text{N}}$)), as well as one defect in ZnO, namely lithium on the zinc
site (ZnO:Li$_{\text{Zn}}$). Experimental identification of defects is often very difficult and frequently
controversial. In order to check our methodology, we wanted to identify benchmark cases where the experimental situation
is clear-cut. GaN:C$_{\text{N}}$ and ZnO:Li$_{\text{Zn}}$ serve this purpose.

Although the focus is on nonradiative transitions, luminescence experiments are frequently used to analyze
rates of the various processes, radiative as well as nonradiative.  There is general consensus that
GaN:C$_{\text{N}}$ gives rise to a yellow luminescence band, \cite{Ogino_JJAP_1980,Seager_JL_2004,Lyons_APL_2010}
and ZnO:Li$_{\text{Zn}}$ to an orange luminescence band.\cite{Meyer_pss_2004,Lyons_2014}
These two bands arise due to the recombination of an electron in the conduction band and a hole
bound to a defect. In both of these cases the acceptor level is in the lower part of the band gap.

Nonradiative hole capture rates for deep acceptors can be determined from luminescence experiments
in the following way.\cite{Reshchikov_PL_2012} In $n$-type samples photo-generated holes are captured
by acceptors in a predominantly nonradiative process (this conclusion stems from the fact that the
resulting capture rates are orders of magnitude higher than possible radiative capture rates,
as discussed in Sec.\ \ref{def}). Subsequently, these captured holes recombine with electrons in
the conduction band, a process believed to be predominantly radiative, giving rise to the aforementioned
luminescence bands.\cite{Reshchikov_PL_2012}

When the temperature is increased, the radiative transition is quenched because captured holes
are re-emitted back into the valence band. Therefore, the measurement of the thermal quenching
of a particular luminescence band as a function of temperature provides information about the
hole emission coefficient $Q_p$. The parameters needed to determine $Q_p$ using this procedure
are the radiative lifetime $\tau_{\text{rad}}=1/C_n n$, measured separately from time-dependent
photoluminescence decay, and the quantum efficiency of the band with respect to all other recombination
channels. \cite{Reshchikov_PL_2012} The hole capture coefficient $C_p$ is determined from $Q_p$ using
the detailed balance equation. \cite{Abakumov,Reshchikov_PL_2012} For acceptors in GaN and
ZnO nonradiative hole capture coefficients determined in this way are summarized in
Ref.\ \onlinecite{Reshchikov_AIP_2014}.

The GaN:(Zn$_{\text{Ga}}$-$V_{\text{N}}$) defect, finally, has been included in order to compare our approach to
that of Ref.~\ \onlinecite{Shi_PRL_2012}, where nonradiative hole capture at this defect was studied.


\subsection{C$_{\text{N}}$ in GaN \label{CN-res}}

\subsubsection{GaN:C$_{\text{N}}$ and yellow luminescence}

Carbon is one of most abundant impurities in GaN, especially if grown by metal organic chemical
vapor deposition, where organic precursors are used. A clear link has been established
\cite{Ogino_JJAP_1980,Seager_JL_2004} between the concentration of carbon and the intensity of a
yellow luminescence (YL) band that peaks at about 2.2 eV. Contrary to earlier suggestions of
carbon being a shallow acceptor, Lyons \emph{et al.}\ have shown, using hybrid density functional
calculations, that carbon on the nitrogen site is in fact a deep acceptor. \cite{Lyons_APL_2010}
Calculations \cite{Lyons_APL_2010,Demchenko_PRL_2013,Lyons_PRB_2014} yield a $(0/-)$ charge-state
transition level $\Delta E$=0.9-1.1 eV above the VBM. In conjunction with a large lattice relaxation
this corresponds to a peak very close to 2.2 eV for the optical transition whereby a neutral defect
captures an electron from the conduction band. Recently we have employed first-principles calculations
to determine effective parameters (average phonon frequencies and the Huang-Rhys factors, see Sec.~\ref{vib-pr}
and Table \ref{Table1}) that describe the shape and temperature dependence of luminescence bands. In the
case of C$_{\text{N}}$ excellent agreement with experimental results \cite{Ogino_JJAP_1980,Reshchikov_AIP_2014}
was demonstrated. \cite{Alkauskas_PRL_2012}

\subsubsection{Configuration coordinate diagram}

\begin{figure}
\includegraphics[width=8.5cm]{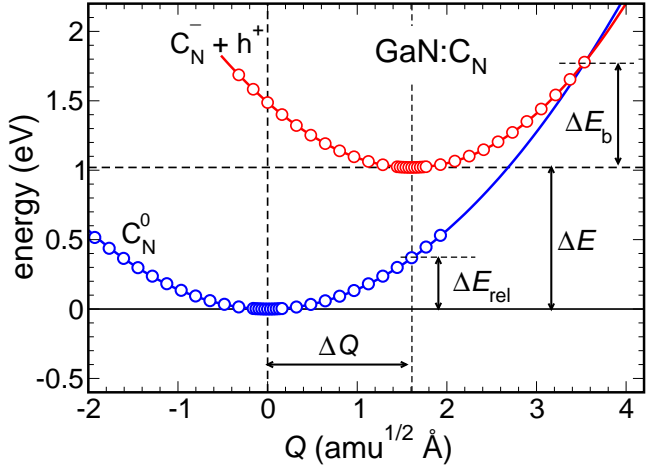}
\caption{(Color online) Calculated 1D cc diagram for hole capture at the C$_\text{N}$ defect in
GaN [process (1) in Fig.\ \ref{SRH}(b)]. Symbols: calculated values; solid line: parabolic fit.
The defect in its negative charge state captures a hole, thus becoming neutral.
$\Delta E$ is the energy difference between the minima of the two potential energy curves,
$\Delta E_{\rm rel}$ is the relaxation energy in the ground state,
$\Delta E_{\rm b}$ is the ``classical'' barrier for the nonradiative process,
and
$\Delta Q$ is the displacement between the two potential energy curves [Eq.\ (\ref{vib2})].}
\label{CN-1D-CCD}
\end{figure}

A 1D cc diagram relevant for the capture of holes at GaN:C$_{\text{N}}$ is shown in Fig.\ \ref{CN-1D-CCD}.
The excited state corresponds to the defect in the negative charge state and a hole in the valence band,
and the ground state corresponds to a neutral charge state. The  configuration coordinate $Q$ was
described in Sec.\ \ref{vib-pr}, and is the same as used in our calculations of luminescence lineshapes.
\cite{Alkauskas_PRL_2012} In the case of GaN:C$_{\text{N}}$,\cite{Lyons_APL_2010} the biggest contribution 
to $\Delta Q$ comes from the the relaxation of C and Ga atoms, which results in the shortening of C-Ga bond 
lengths by $0.05-0.09$ \AA, as the charge state changes from neutral to $-1$. A smaller
contribution comes from the relaxation of next-nearest N atoms, which results in a slight increase of C-N distances, 
expected from electrostatic arguments. Potential energy surfaces in the two charge states were mapped along this
configuration coordinate. The separation between the minima of the two potential energy surfaces
$\Delta E$ corresponds to the energy of the $(0/-)$ charge-state transition level with respect to
the VBM. Our calculated value for $\Delta E$=1.02 eV is slightly larger than the one reported
in Ref.~ \onlinecite{Lyons_APL_2010} due to more stringent convergence criteria. The minima of the 
two potential energy surfaces are offset  horizontally by $\Delta Q$=1.61 amu$^{1/2}$\AA$ $ [Eq.\ (\ref{vib2})].
An important parameter is the relaxation energy in the ground state $\Delta E_{\rm rel}$ [Fig.\ \ref{CN-1D-CCD}].
For $C_{\text{N}}^0$ calculations yield $\Delta E_{\rm rel}$=0.37 eV.

The two potential energy curves intersect at $\Delta E_b$=0.73 eV above the minimum of the
excited state. We might thus expect that the nonradiative carrier capture is a temperature-activated
process, since the coupling between two potential energy surfaces is always most efficient close
to the crossing point.\cite{DiBartolo}

\subsubsection{Calculated hole capture coefficients}

The real situation is not so straightforward because of the occurrence of two competing factors.
On the one hand, when the temperature is raised, higher-lying vibrational levels $\chi_{im}$
of the excited electronic state  [see Eqs.~ (\ref{F2}) and (\ref{r5})] become populated. Vibrational levels
that are closer in energy to the crossing point of the two potential energy curves
yield larger contributions to the overall rate. Thus, if this was the only factor, the
nonradiative capture rate for GaN:C$_{\text{N}}$ would increase as a function of temperature.
On the other hand, however, the scaling factor $f(T)$ decreases with temperature, as per
Eq.\ (\ref{temp-scaling}), because an increasingly faster hole has less chance of being
captured by a negative acceptor.

\begin{figure}
\includegraphics[width=8.5cm]{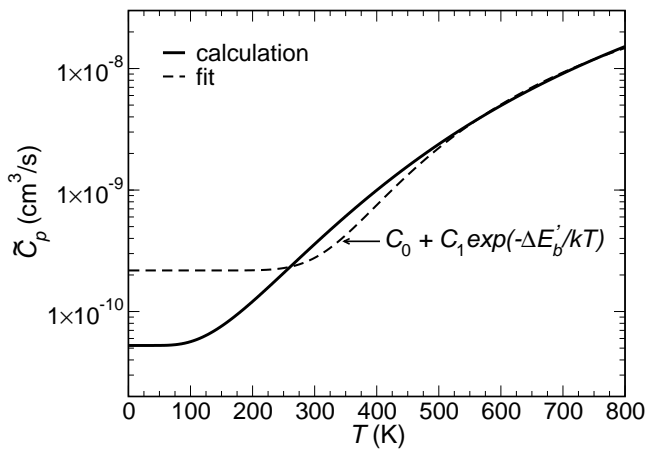}
\caption{``Nominal'' nonradiative hole capture rate $\tilde{C}_{p}$ [Eq.~(\ref{r5})]
at the C$_{\text{N}}$ defect in GaN as a function of temperature. Solid line: calculations;
dotted line: a fit according to Eq.\ (\ref{fit}).
}
\label{C-nonrad-test}
\end{figure}

It is instructive to consider the first effect separately. In Fig.\ \ref{C-nonrad-test} we show
the calculated nonradiative hole capture rate if the second effect is completely neglected, i.e.,
for $f=1$. This is the capture rate $\tilde{C}_p$ that is discussed in Section \ref{scatter}.
The process is indeed temperature-activated. At high temperatures, the dependence is often fitted
to a function of the form
\begin{equation}
\tilde{C}_p(T)=C_0+C_1 \text{exp}(-\Delta E'_{\rm b}/kT)
\label{fit}
\end{equation}
with a temperature-independent part and a temperature-activated part. The use of such a form
is at the core of the famous Mott-Seitz formula for temperature quenching of luminescence bands.
\cite{DiBartolo} The fit is shown in Fig.\ \ref{C-nonrad-test}. From the fit one can derive an
effective barrier $\Delta E'_{\rm b}$=0.23 eV, which is significantly smaller than the ``classical''
barrier $\Delta E_{\rm b}$=0.73 eV. This is a typical result and happens because of the quantum-mechanical
tunneling \cite{DiBartolo} that is considered in the quantum treatment but absent in a classical
description.

\begin{figure}
\includegraphics[width=8.5cm]{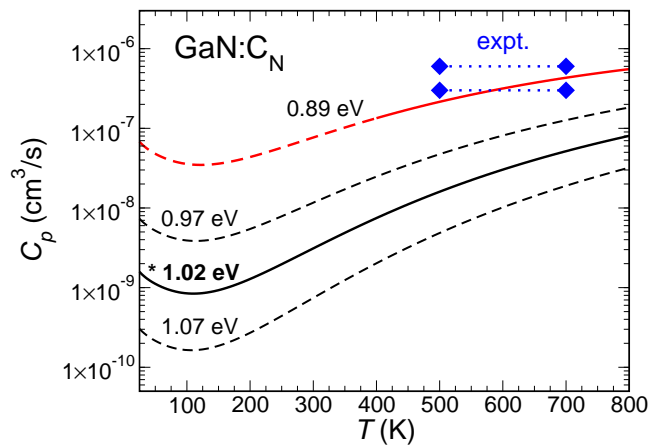}
\caption{(Color online) Black solid line: calculated nonradiative hole capture rate $C_p$ at the
C$_{\text{N}}$ defect in GaN, for $\Delta E$=1.02 eV, as determined from our first-principles
calculations. Black dashed lines: the same for $\Delta E$=1.02$\pm$0.05 eV. Red solid and dotted
line: the same, but for $\Delta E$=0.89 eV, modeling the decrease of $\Delta E$  at high
temperatures. Blue dotted horizontal lines and diamonds: experimental data from Ref.~\
\onlinecite{Reshchikov_AIP_2014}; the latter were determined for several different samples in
the temperature range 500-700 K.
}
\label{C-nonrad}
\end{figure}

The actual hole capture coefficient, including the scaling factor [Eq.\ (\ref{temp-scaling})],
is presented in Fig.\ \ref{C-nonrad} (solid black line). At $T$=300 K, our calculated
value is $C_p=3.1\times10^{-9}$ cm$^{3}$s$^{-1}$. To determine the sensitivity of the final
result on the parameters of our calculation, we have also determined the capture coefficient
for $\Delta E$=1.02$\pm$0.05 eV (Fig.\ \ref{C-nonrad}). A change in $\Delta E$ by just 0.05 eV
translates into a change of $C_p$ by a factor of $\sim3-4$ (black dashed lines in Fig.\ \ref{C-nonrad}).
Since $\Delta E_{\rm rel} < \Delta E$ (cf.\ Fig.\ \ref{CN-1D-CCD}), larger values of $\Delta E$ yield
larger barriers $\Delta E_b$ and therefore smaller capture coefficients $C_p$. The strong dependence
of nonradiative transitions on $\Delta E$ is well documented,\cite{DiBartolo} and was recently
emphasized again in Ref.~\ \onlinecite{Shi_PRL_2012}. This sensitivity stems from the temperature-activated
behavior discussed in the preceding paragraphs. We can consider the range of capture coefficients
$C_p$ shown in Fig.~\ \ref{C-nonrad} to represent a theoretical ``error bar'' regarding
the 0 K value of $\Delta E$ that is used in the calculation.

\subsubsection{Comparison with experiment}

In Refs.~\ \onlinecite{Reshchikov_AIP_2014} and $ $ \onlinecite{Demchenko_PRL_2013} the
hole capture coefficient $C_p$ for C$_{\rm N}$ was determined in the temperature range 500-700 K,
at which quenching of the luminescence occurs. $C_p$ was assumed to be weakly dependent on temperature in
this range, and the values obtained for different samples were $C_p=3\times10^{-7}$ and
$6\times10^{-7}$ cm$^{3}$s$^{-1}$.\cite{Reshchikov_AIP_2014} These results are shown by horizontal dotted
lines and diamonds in Fig.\ \ref{C-nonrad}. At $T$=600 K, i.e., the midpoint of the 500-700 K temperature range
where the quenching occurs, our calculated value for $\Delta E=1.02$ eV is $C_p=3.1\times10^{-8}$ cm$^{3}$s$^{-1}$,
i.e., about an order of magnitude smaller than the experimental result. The corresponding values for $\Delta E$=0.97 eV
and 1.07 eV are $8.0\times10^{-8}$ and $1.0\times10^{-8}$ cm$^{3}$s$^{-1}$ (cf.\ Fig.\ \ref{C-nonrad}).
Thus, variations of $\Delta E$ by 0.05 eV do not remove the difference between theory and experiment.

The apparent discrepancy between experiment and theory can be explained as follows.
The calculated $\Delta E$ of 1.02 eV corresponds to the ionization potential of the C$_{\text{N}}$ acceptor at $T$=0 K,
while the comparison with experiment is made for $T$$\approx$600 K. At 600 K the bulk band gap of GaN
shrinks from its $T$=0 K value of 3.50 eV to about 3.24 eV. \cite{Nepal_APL_2005} This decrease of
the band gap will affect the ionization potential of the acceptor. We are not in a position to address
this fully from first principles, but we can obtain a zero-order estimate of the effect of
temperature on $\Delta E$ by assuming that (i) the $(0/-)$ charge-state transition level remains
the same on the absolute energy scale when the band gap changes,\cite{Alkauskas_PRL_2008,
Lyons_PRB_2014,Chen_PRB_2013} and (ii) the VBM and the conduction-band minimum are equally affected,
i.e., that as a function of temperature they move symmetrically on the same absolute energy scale. Based on
these assumptions we estimate a decrease of the ionization potential by about $0.13$ eV in going from 0 K to 600 K.
In Fig.\ \ref{C-nonrad} we have included a curve for $C_p$ as a function of $T$ for $\Delta E$=0.89 eV
(red solid and dashed curve). This curve is physically meaningful only at temperatures around 600 K, corresponding
to a significantly reduced value of $\Delta E$. At $T$=600 K, the calculated value $C_p=3.1\times10^{-7}$
cm$^{3}$s$^{-1}$ is in excellent agreement with the experimental data. \cite{Reshchikov_AIP_2014,Demchenko_PRL_2013}

We thus find that significant variations of the calculated $C_p$ can result from the temperature
dependence of $\Delta E$.  For comparison with experiments carried out at elevated temperatures
this can make a difference in $C_p$ of about an order of magnitude. This dependence has not been
studied in the past, neither theoretically nor experimentally. Our findings indicate that this
will be a fruitful area of future work on nonradiative carrier capture. However, even in the absence
of a rigorous analysis of the temperature dependence, we can conclude that our calculated values
of $C_p$ are in very good agreement with experimental data.


\subsection{Li$_{\text{Zn}}$ in ZnO \label{Li}}

\subsubsection{ZnO:Li$_{\text{Zn}}$ and orange luminescence}

Li$_{\text{Zn}}$ in ZnO is one of the most studied defects in ZnO. While it was initially hoped
that Li$_{\text{Zn}}$ might be a shallow acceptor leading to $p$-type doping, it is now clear
that this defect is a very deep acceptor. Meyer \emph{et al.} suggested that Li$_{\text{Zn}}$
gives rise to a broad orange luminescence (OL) band peaking at about 2.1 eV. \cite{Meyer_pss_2004}
The ionization energy was deduced to be at least 0.5 eV. The analysis based on the thermal quenching of
luminescence lines confirms this and yields values for the ionization potential $\Delta E=0.46-0.55$ eV;
\cite{Reshchikov_PB_2007,Reshchikov_AIP_2014} the different values are for different ZnO samples.
Recent theoretical work based on the application of the generalized Koopman's theorem \cite{Lany_PRB_2009} and on
hybrid density functionals \cite{Du_PRB_2009,Carvalho_PRB_2009,Bjorheim_JPCC_2012,Lyons_2014} has confirmed
that Li$_{\text{Zn}}$ is indeed a deep acceptor with a ionization energy $>$0.3 eV.
The neutral charge state of the defect corresponds to a small polaron bound to an oxygen atom that is
adjacent to Li.

\subsubsection{Configuration coordinate diagram}

\begin{figure}
\includegraphics[width=8.5cm]{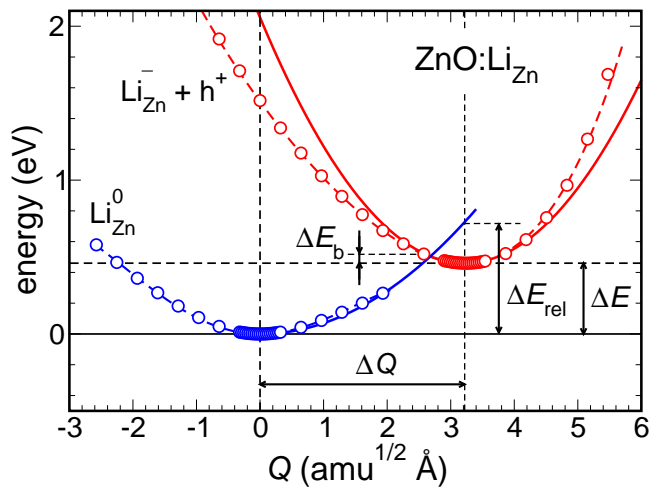}
\caption{(Color online) Calculated 1D cc diagram for hole capture at the Li$_\text{Zn}$ defect in ZnO.
Physical quantities as in Fig.\ \ref{CN-1D-CCD}. Symbols: calculated values;
solid lines: parabolic fit; dashed lines: a fit to a fourth-order polynomial.
Note that for the neutral charge state the parabolic fit is performed only for $Q>0$.
}
\label{Li-1D-CCD}
\end{figure}

The calculated 1D cc diagram relevant for hole capture by a Li$_{\text{Zn}}$ defect in ZnO is shown in Fig.\
\ref{Li-1D-CCD}. The calculations were consistently performed with the HSE hybrid functional (mixing
parameter $\alpha$=$0.38$) and resulted in a value $\Delta E$=0.46 eV. \cite{Lyons_2014}
In the negatively charged state the defect posses a C$_{3v}$ symmetry, with the axial Li-O
bond-length of 2.00 \AA, being only slightly larger than azimuthal Li-O bond-lenghts of
1.96 \AA. In the neutral charge state the Li atom undergoes a huge relaxation 
\cite{Du_PRB_2009,Carvalho_PRB_2009,Bjorheim_JPCC_2012,Lyons_2014}
of about 0.4 \AA. This results in the increase of one of the Li-O bond-lengths to 2.61 \AA,
and the decrease of the remaining three to about 1.87 \AA. The hole is bound to the oxygen atom
involved in the long bond;\cite{Carvalho_PRB_2009,Bjorheim_JPCC_2012,Lyons_2014} three Zn atoms 
closest to this oxygen relax outwards by about 0.16 \AA.

Note, that in contrast to GaN:C$_{\text{N}}$, in the case of ZnO:Li$_{\text{Zn}}$
$\Delta E_{\rm rel} > \Delta E$. Therefore, the two potential energy curves intersect for $Q < \Delta Q$.
Another difference with GaN:C$_\text{N}$ is that the potential energy curves for ZnO:Li$_{\text{Zn}}$
are very anharmonic.  The solid curves in Fig.\ \ref{Li-1D-CCD} present parabolic fits to the potential
energy values, while dashed lines are fits to a fourth-order polynomial. We need to make an important
point here: while we use the harmonic approximation in the present work, there is no requirement that
the parabolic fit to the potential energy surface be performed for the entire range of $Q$ values.
Indeed, in order to capture the most essential physics we should focus on those $Q$ values that are
relevant for the transitions under investigation, i.e., the range of $Q$ values where the
potential energy curves cross. Therefore, in Fig.\ \ref{Li-1D-CCD}, the potential energy
curve for the neutral charge state has been fitted to a parabolic curve only for $Q>0$.
The effective phonon frequencies derived from the parabolic fits are included in Table \ref{Table1}.

\subsubsection{Calculated hole capture coefficients and comparison with experiment}

\begin{figure}
\includegraphics[width=8.5cm]{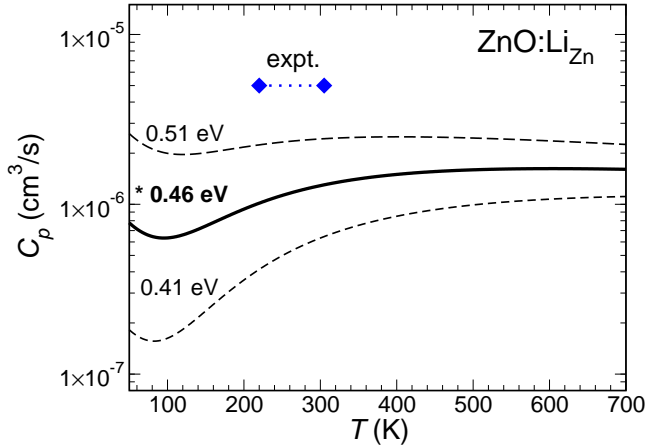}
\caption{(Color online) Solid black line: calculated nonradiative hole capture coefficient $C_p$ at the Li$_{\text{Zn}}$ defect
in ZnO, for $\Delta E$=0.46 eV, as determined from our first-principles calculations. Black dashed lines: the same,
but for $\Delta E$=0.46$\pm$0.05 eV. Blue dashed lines and diamonds: experimental data from Ref.~\
\onlinecite{Reshchikov_PB_2007}.
}
\label{Li-nonrad}
\end{figure}

The calculated hole capture coefficient $C_p$ for Li$_\text{Zn}$ in ZnO is shown in Fig.\ \ref{Li-nonrad}
(black solid curve). Our room-temperature value is $C_p$=1.3$\times10^{-6}$ cm$^{3}$s$^{-1}$.
In Ref.~\ \onlinecite{Reshchikov_AIP_2014} the capture coefficient was determined from the quenching of the OL.
The quenching occurred in the temperature range 225-300 K, and fitting yielded values $C_p\approx5\times10^{-6}$
cm$^{3}$s$^{-1}$, as indicated in Fig.\ \ref{Li-nonrad} (diamonds and horizontal dashed line).
To determine the sensitivity of the final result on the value
of $\Delta E$, we have calculated $C_p$ for $\Delta E$=0.46$\pm$0.05 eV. As seen in Fig.\ \ref{Li-nonrad},
changes in $\Delta E$ by 0.05 eV result in changes in $C_p$ by about a factor of 2.
However, in contrast to GaN:C$_{\text{N}}$, smaller $\Delta E$ yield larger $C_p$ values. This can be understood
by considering the 1D cc diagram in Fig.\ \ref{Li-1D-CCD}. Since $\Delta E_{\rm rel} < \Delta E$ for all three values of
$\Delta E$, larger $\Delta E$ yield smaller barriers $\Delta E_b$, and thus larger capture coefficients $C_p$.

The temperature dependence of $\Delta E$ that was important for GaN:C$_{\text{N}}$ at 600 K, is not
substantial for ZnO:Li$_{\text{Zn}}$ at 300 K. At 300 K the band gap of ZnO is lower by about 0.03 eV
compared to its 0 K value. Considerations similar to the one performed for GaN:C$_{\text{N}}$ would yield
a change in $\Delta E$ by 0.015 eV, translating into a change of $C_p$ by at most 10\%.

Both theory and experiment thus confirm that the coefficient $C_p$ for ZnO:Li$_\text{Zn}$ is larger than
that for GaN:C$_{\text{N}}$ by about an order of magnitude. The main reason is the fact that the potential
energy curves for ZnO:Li$_\text{Zn}$ intersect close to the minimum of the excited state (Fig. \ref{Li-1D-CCD}),
rendering the nonradiative process more likely even at low temperatures. In addition, we find that the
temperature dependence of $C_p$ is significantly weaker for ZnO:Li$_\text{Zn}$ than in the case of GaN:C$_{\text{N}}$.
This is because the ``classical'' barrier $\Delta E_{\rm b}$ for the nonradiative transition is very small for
ZnO:Li$_{\text{Zn}}$ (Fig.\ \ref{Li-1D-CCD}).

Overall, we can again conclude that first-principles calculations of hole capture coefficients
at ZnO:Li$_{\text{Zn}}$ agree very favorably with experimental data.


\subsection{Zn$_{\text{Ga}}$-$V_{\text{N}}$ in GaN \label{ZnGa-VN}}

\subsubsection{Defect properties and configuration coordinate diagram}

To compare our methodology with the approach used in Ref.~ \onlinecite{Shi_PRL_2012}, we have also
studied hole capture by a neutral Zn$_{\text{Ga}}$-$V_{\text{N}}$ complex in GaN, the example
that was studied in that work. In contrast to GaN:C$_{\text{N}}$ and ZnO:Li$_{\text{Zn}}$,
GaN:(Zn$_{\text{Ga}}$-$V_{\text{N}}$) is a deep donor with $(+2/+)$ and $(+/0)$ charge-state transition
levels in the lower part of the band gap. Its defect wavefunction is derived mostly from Ga states.
To the best of our knowledge, no direct experimental data is available for this
defect.

The approach of Shi and Wang is based on the adiabatic approach to nonradiative transitions, employing
the formula derived by Freed and Jortner. \cite{Freed_JCP_1970} This should be contrasted to the static
approach used in the current work. The distinction between the two approaches will be discussed in Section
\ref{st-ad}. The calculation of Ref.~ \onlinecite{Shi_PRL_2012} includes the coupling to all phonon modes.

\begin{figure}
\includegraphics[width=8.5cm]{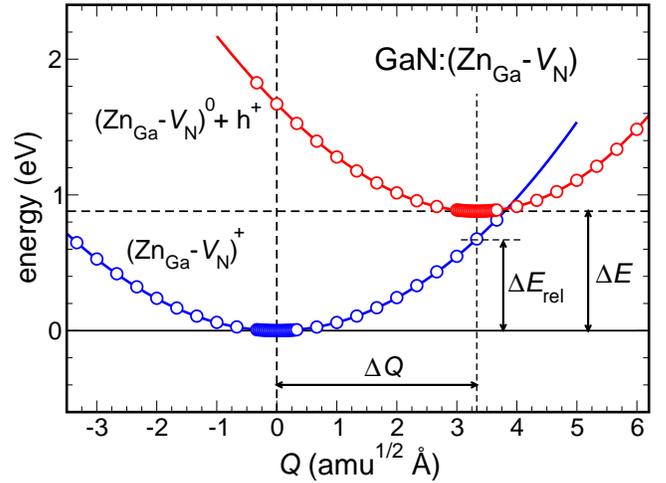}
\caption{(Color online) 1D configuration coordinate diagram for hole capture at the Zn$_{\text{Ga}}$-$V_{\text{N}}$
defect in GaN. Physical quantities as in Fig.\ \ref{CN-1D-CCD}. Symbols: calculated values; solid lines:
parabolic fit.
}
\label{ZnGa-VN-1D-CCD}
\end{figure}

A 1D cc diagram for Zn$_{\text{Ga}}$-$V_{\text{N}}$, relevant for hole capture by a neutral center,
is shown in Fig.\ \ref{ZnGa-VN-1D-CCD}.  The values are again calculated with a hybrid functional with
$\alpha=0.31$. With respect to the neutral charge state, four cation atoms surrounding the vacancy 
experience and outward relaxation in the positive chare state: Ga atoms relax by about 0.20 \AA,
while the Zn atom relaxes by about 0.12 \AA. A relaxation away from the vacancy in the positive
charge state is consistent with electrostatic arguments.

Our calculated value of $\Delta E$=0.88 eV, corresponding to the (0/+) charge-state transition
level, was used in Fig.~ \ref{ZnGa-VN-1D-CCD}. Our calculated $\Delta E$ value is close to the value $\Delta E$=0.90 eV 
of Shi and Wang, \cite{Shi_PRL_2012} but this agreement is to some extent accidental, since the details of the calculations
differ: Shi and Wang's value was determined based on a supercell of 300 atoms, without any finite-size correction,
using HSE calculations with $\alpha=0.25$ but based on atomic geometries determined at the PBE level.
We estimate that their value after inclusion of finite-size corrections would be 0.82 eV.
As a check we performed the calculations using the hybrid functional with $\alpha=0.25$.
Using the geometry obtained at the HSE level we obtain a value of $0.67$ eV for the $(0/+)$ charge-state transition level.
If instead the geometry is obtained at the PBE level, the value is $0.64$ eV.

\subsubsection{Hole capture coefficients}

\begin{figure}
\includegraphics[width=8.5cm]{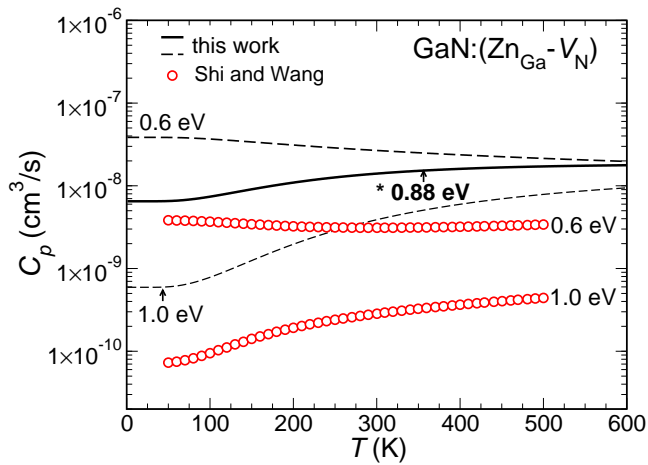}
\caption{(Color online) Nonradiative hole capture rate at the  Zn$_{\text{Ga}}$-$V_{\text{N}}$ defect in GaN for
different values of $\Delta E$.
Solid line and dashed lines: this work; symbols: results from Ref.~\ \onlinecite{Shi_PRL_2012}.
}
\label{ZnGa-VN-nonrad}
\end{figure}

In Fig.\ \ref{ZnGa-VN-nonrad} we compare the results of Shi and Wang \cite{Shi_PRL_2012}
with our present results. Shi and Wang's values of $C_p$ are presented as a function of $T$
for two different values of $\Delta E$, namely 0.6 and 1.0 eV, that encompass
the theoretical values quoted above. The two sets of calculations agree quite well with each other
if we focus on the temperature dependence of $C_p$ and the trends as a function of $\Delta E$.
However, our calculated room-temperature values for $C_p$, namely $2.4 \times 10^{-9}$-$2.9 \times 10^{-8}$ cm$^{3}$s$^{-1}$
are consistently about an order of magnitude larger than those of Shi and Wang, $C_p$=$2.6\times$10$^{-10}$-$2.9
\times10^{-9}$ cm$^{3}$s$^{-1}$ for values of $\Delta E$ that range between 0.6 and 1.0 eV. Also, our first-principles
value for the room-temperature capture coefficient for this defect, obtained for $\Delta E = 0.88$ eV (black solid line
in Fig.\ \ref{ZnGa-VN-nonrad}), $C_p=1.0\times10^{-8}$ cm$^{3}$s$^{-1}$, is about 16 times larger than the corresponding
value of Shi and Wang.\cite{Shi_PRL_2012}

It is important to try and identify the origin of these discrepancies, in order to assess whether they
are due to differences in the computational approach and/or differences in the methodology.  This type
of analysis helps in establishing the validity and reliability of the overall approach
and provides important insights. One of the differences between the two calculations is the treatment of
the electronic structure of the defect. In our approach all the parameters have been determined
consistently using the hybrid functional with $\alpha$=$0.31$. In Ref.~\ \onlinecite{Shi_PRL_2012} most of the
results, such as the ground-state geometries, phonon spectra, and electron-phonon coupling matrix elements,
were determined using the semilocal PBE functional. The two sets of calculations yield, in particular,
very different relaxation energies $\Delta E_{\rm rel}$ for the ground state. In our approach, this
relaxation energy is 0.67 eV, while the value of 0.43 eV was used in Ref.~\ \onlinecite{Shi_PRL_2012}.
Unfortunately, we were not able to reproduce this result. Still, to gain insight into
the sensitivity to the underlying electronic structure we repeated our
calculations using a hybrid functional with a fraction $\alpha=0.25$ instead of $\alpha=0.31$.
This yielded $\Delta E_{\rm rel}=0.62$ eV. The calculated values of $C_p$ decrease by up to a factor of
2, bringing our results in better agreement with Ref.~\ \onlinecite{Shi_PRL_2012}.


Another difference between our approach and that of Shi and Wang \cite{Shi_PRL_2012}
is the inclusion of the scaling factor $f$, discussed in Sec.\ \ref{scatter}.
In the case of the Zn$_{\text{Ga}}$-$V_{\text{N}}$ complex in GaN, the coupling between the hole state
at the $\Gamma$ point and the defect state is evaluated for a positively charged supercell.
If we consider a particle with zero kinetic energy interacting with a repulsive potential, the
particle would be repelled to infinity. Therefore the matrix element $\tilde{W}_{if}$ as well as
the product $\tilde{W}_{if}^2\tilde{V}$ will tend to zero as a function of increasing
supercell size [cf.\ Eqs.\ (\ref{e-ph-tilde}), (\ref{e-ph-tilde1}), and (\ref{sca3})].
As a result, performing the calculation of electron-phonon coupling for
increasingly larger supercells will not lead to converged results, but rather result in a
decreasing value of the coefficient $C_p$. It is to combat such effects that the scaling factor $f$
[Eq.~(\ref{sca4})] has been introduced in the first place.\cite{Paessler_pssb_1976,Bonch-Bruevich_FTT_1959}
This effect was not considered in Ref.~\ \onlinecite{Shi_PRL_2012} and hence their values obtained
for a 300-atom supercell are underestimated.  Using the information provided in Ref.~\ \onlinecite{Shi_PRL_2012},
we estimate that the values of $C_p$ in Ref.\ \onlinecite{Shi_PRL_2012} are probably too small by a
factor of about 1.5. Proper inclusion  of the scaling factor $f$ would bring the values of Shi and Wang
\cite{Shi_PRL_2012} closer to ours.

The two sources of discrepancies considered so far still do not account for the fact that
our results are more than one order of magnitude larger than those of Shi and Wang.\cite{Shi_PRL_2012}
One might argue that the consideration of all phonon degrees of freedom in Ref.~ \onlinecite{Shi_PRL_2012}
is in principle more accurate than the reduction of the problem to a single effective phonon frequency,
as we do in the present work. However, the analysis we will present in Sec.\ \ref{multi-1D} indicates
that one should expect the 1D model to {\it underestimate} the true result.
This consideration therefore does not resolve the discrepancy either. More careful scrutiny traces the
difference between the two approaches to the core assumptions of the method employed in
Ref.~\ \onlinecite{Shi_PRL_2012}, which was originally proposed by Freed and Jortner.\cite{Freed_JCP_1970}
The formula of Freed and Jortner is derived from the so-called adiabatic approach
within the Condon approximation.\cite{Huang_SS_1981,Peuker_pssb_1982} This approach has been recognized
to underestimate nonradiative capture rates,\cite{Huang_SS_1981,Peuker_pssb_1982} an issue to be addressed
in Sec.\ \ref{st-ad}.


\section{Discussion \label{disc}}
In this Section we critically analyze our theoretical approach and compare calculated capture coefficients
to those in other materials.

\subsection{The strength of electron-phonon coupling \label{e-ph7}}

In this section we discuss the strength of electron-phonon coupling at defects considered
in this work, as expressed by the matrix elements $\tilde{W}_{if}$ (Table \ref{Table1}).
It is informative to estimate the maximum possible value of this matrix element, for typical
values of $\Delta E$ and $\Delta Q$.  According to Eq.\ (\ref{e-ph8}), we have to find the maximum
value of $\Braket {\tilde{\psi}_{i} | \partial \psi_{f}/\partial Q} $, where $\tilde{\psi}_i$ is the perturbed valence
band and $\psi_f$ is the defect state. Let us assume that for $Q$=$\Delta Q$ ($Q$=$0$ corresponds
to the equilibrium configuration of the ground state) the defect state acquires completely the
character of the perturbed valence band, albeit still localized on $M_d$ atoms, with the total
number of atoms in the supercell being $M_b$. We can then use our knowledge about the degree of
localization of $\tilde{\psi}_{i}$ and $\psi_{f}$, along with the normalization of the wavefunctions, to estimate
the matrix element.  Replacing the derivative by its finite-difference expression, we see that the maximum
value of $\Braket {\tilde{\psi}_{i} | \partial \psi_{f}/\partial Q} $ is $\sim\sqrt{M_d/M_b}/\Delta Q$.  Therefore
a large $\tilde{W}_{if}$ would correspond to $\tilde{W}_{if}\approx\Delta E \sqrt{M_d/M_b}/\Delta Q$ as per
Eq.\ (\ref{e-ph8}).

Taking typical values for $\Delta E$ and $\Delta Q$ from Table \ref{Table1}, assuming that
$M_d$$\sim$4, $M_b$=96, and also that coupling to only one valence band is present,
we obtain $\tilde{W}_{if}\sim1-7\times10^{-2}$ eV/(amu$^{1/2}$\AA). As shown in
Table \ref{Table1}, the actual values for the defects considered here are all within this
range. Our estimate of the maximum value of $\tilde{W}_{if}$ yields important insight:
for the defects considered, electron-phonon coupling pertaining to the special
mode $Q$ from Eq.\ (\ref{vib1}) is, in fact, very strong. In the case of hole
capture studied in this work, the coupling is very effective for the acceptor
defects with electronic states derived from anion 2$p$ orbitals (GaN:C$_{\text{N}}$
and ZnO:Li$_{\text{Zn}}$), but is also quite significant for the donor defect
[GaN:(Zn$_{\text{Ga}}$-$V_{\text{N}}$)] with electronic states derived
mainly from the cation (Ga) 3$s$ states.

\subsection{1D vs.\ multi-dimensional treatment \label{multi-1D}}

The nonradiative capture processes studied in this work have been analyzed employing
1D configuration coordinate diagrams. Here, we critically review the range
of applicability of this approach.

It is known that different phonons have different functions during a nonradiative transition.
\cite{Stoneham_RPP_1981,Freed_JCP_1970} Vibrational modes that couple the two electronic states very efficiently,
i.e., those that yield large electron-phonon matrix elements, are called the ``promoting'' modes, whereas the
modes that couple strongest to the distortion of the geometry during the transition are called the ``accepting''
modes.\cite{Stoneham_RPP_1981} A particular mode can, of course, be both ``accepting'' and ``promoting''.

The 1D treatment of Sec.\ \ref{vib-pr} essentially considers only the ``accepting'' modes, reducing the treatment
to one effective mode that has the strongest possible ``accepting'' character, being completely parallel to the
distortion of the defect geometry during the transition [Eq.\ (\ref{vib1})]. For the nonradiative transition to
be effective this mode must also have a lot of ``promoting'' character, i.e., it must couple the two electronic
states by producing a sizable electron-phonon matrix element $W_{if}$ (or $\tilde{W}_{if}$). One therefore expects
the calculated transition rate in the 1D treatment to be somewhat smaller than the real one; i.e., the 1D approximation
should provide a lower bound.

As we have shown in Sec.\ \ref{e-ph7}, for all the defects studied in this work the ``accepting'' mode is
characterized by a large electron-phonon matrix element, and therefore has a lot of ``promoting'' character.
This justifies the use of the 1D approach for such defects.  Still, it should be acknowledged that our calculated
capture coefficients might be slightly underestimated. In order to include all phonon modes in a rigorous
theoretical treatment, an efficient algorithm to calculate electron-phonon coupling matrix elements for hybrid
functionals is urgently needed. While the method used in the present work [cf. Eq.~(\ref{e-ph8}) and Fig.\ \ref{e-ph4}]
is one algorithm to determine those elements, it is computationally too demanding if all phonons need
to be included.

\subsection{Static and adiabatic formulation of nonradiative transitions \label{st-ad}}

The first-principles methodology presented in this work is based on the static-coupling
approach.\cite{Paessler_pssb_1975,Paessler_CJP_1982,Goguenheim_JAP_1990}
In Sec.\ \ref{der} we stated that this approach is applicable when carrier
capture rates are much smaller than phonon lifetimes and periods of lattice vibrations.\cite{Paessler_pssb_1975}
Now we are in a position to verify this assumption. Note that as per Eq.\ (\ref{rate}), carrier capture
rate can mean two separate things. For a given hole its capture time by any acceptor
of the type $A$ is $\tau_p=1/(C_pN_{A}^{-})$ (measured in s). However, for a given
acceptor in its negative charge state the time it takes to capture a hole is $\tau_A=1/(C_pp)$.
It is the latter quantity that is of importance when determining the range of validity of the static
approach.

For GaN:C$_{\text{N}}$ and ZnO:Li$_{\text{Zn}}$ we compared calculated values to experimental
results summarized in Ref.~ \onlinecite{Reshchikov_AIP_2014}. In these experiments typical hole densities
were $p\approx10^{13}$ cm$^{-3}$.\cite{Reshchikov_JAP_2005} Using the calculated
values for $C_p$ (cf.~Figs.~\ref{C-nonrad}, \ref{Li-nonrad}, \ref{ZnGa-VN-nonrad}),
we estimate hole capture times $\tau_A\approx10^{-4}-10^{-7}$ s. These values
are much larger than typical periods of lattice vibrations $2\pi/\Omega\approx10^{-13}$ s (0.1 ps),
or phonon lifetimes, which are at most a few ps. \cite{Matulionis_APL_2009} Thus, the transition between the two
electronic states [process (1) in Fig.\ \ref{SRH} (b)] is indeed the time-limiting step
in carrier capture. This transition is a rare event; once it happens, the emission of phonons
due to phonon-phonon interaction [process (2) in Fig.\ \ref{SRH}] occurs almost instantaneously.
The range of validity of the static approach can be judged for different defects separately. For
example, in the case of hole capture by GaN:C$_{\text{N}}$ at room temperature, we estimate that the static
approach is valid for hole densities up to $p\sim10^{19}-10^{20}$ cm$^{-3}$.

In parallel with the static approach, which we advocate and justified above, a lot of theoretical work
on nonradiative carrier capture employed the so-called adiabatic coupling scheme.\cite{Huang_PRCL_1950,Freed_JCP_1970}
In this approach one chooses Born-Oppenheimer wavefunctions $\Psi(\{Q\},\{x\})\chi{\{Q\}}$ as a starting point for
perturbation theory. This choice should be contrasted to the choice made in the static approach, namely $\Psi(\{Q_0\},\{x\})\chi{\{Q\}}$,
where $\{Q_0\}$ is a specific fixed atomic configuration (see Sec.~\ref{der}). Many early formulations
based on the adiabatic approach also assumed that electron-phonon matrix elements were independent
of $\{Q\}$,\cite{Huang_PRCL_1950,Freed_JCP_1970} which is the Condon approximation for nonradiative transitions.
However, already in the 1970s it was noticed that the adiabatic approach within the Condon
approximation yielded capture rates significantly smaller than the static approach applied to the same
system.\cite{Paessler_pssb_1975} This result was very surprising at the time; indeed, it was expected that far
from the intersection of potential energy surfaces the two descriptions should yield very similar physics.

The issue was resolved in the early 1980s by Huang, \cite{Huang_SS_1981}
Gutsche and co-workers, \cite{Gutsche_pssb_1982,Peuker_pssb_1982} and Burt.\cite{Burt_JPC_1983}
The discrepancy was found to originate in an inconsistent application of the adiabatic approach.
These authors demonstrated that within the leading order the adiabatic approach {\it does} give
the same answer as the static approach, provided the adiabatic approach is applied consistently.
This can be achieved in one of two ways: (i) by going beyond the Condon approximation in the adiabatic
treatment \cite{Huang_SS_1981}; or (ii) by considering all non-diagonal terms in the
Hamiltonian.\cite{Peuker_pssb_1982} In particular, erroneous omission of non-diagonal terms can
lead to significantly smaller values of capture rates in the adiabatic approach.\cite{Peuker_pssb_1982}
This was exactly the problem of prior theoretical treatments that were based on the adiabatic approach
together with the Condon approximation.\cite{Huang_PRCL_1950,Freed_JCP_1970}

The discussion up to this point applies to the case where nonradiative transitions occur far from the
intersection of two potential energy surfaces. However, when transitions close to the crossing point are
important (as is the case for ZnO:Li$_{\rm Zn}$ and GaN:C$_{\rm N}$, cf.~Figs.~\ref{Li-1D-CCD} and
\ref{ZnGa-VN-1D-CCD}) the adiabatic approach within the Condon approximation fails altogether. In the
adiabatic approach, an avoided crossing  occurs between the two potential energy surfaces, leading to
a strong variation of electronic wavefunctions and making potential energy surfaces very anharmonic.
In such a situation, adiabatic wavefunctions are probably not a good starting point for perturbation
theory.\cite{Paessler_pssb_1975} The static coupling scheme, on the other hand, is still applicable in
the regime close to the intersection of the potential energy surfaces.\cite{Paessler_pssb_1975,Paessler_CJP_1982}

In our opinion, the application of the adiabatic coupling scheme in Ref.~ \onlinecite{Shi_PRL_2012},
versus the static scheme employed in the present work, is the major reason for the differences between
our results and those of Ref.~ \onlinecite{Shi_PRL_2012} (Fig.\ \ref{ZnGa-VN-nonrad}).
The results of Ref.~ \onlinecite{Shi_PRL_2012} were obtained based on the formula of
Freed and Jortner, \cite{Freed_JCP_1970} which uses the adiabatic coupling scheme within the Condon
approximation and will thus tend to underestimate the values for capture coefficients.
\cite{Huang_SS_1981,Gutsche_pssb_1982,Peuker_pssb_1982,Burt_JPC_1983}

The theoretical foundations of nonradiative capture due to multi-phonon emission were laid in
Refs.~ \onlinecite{Huang_PRCL_1950, Kubo_PTP_1955,Gummel_AP_1957,Henry_PRB_1977,Freed_JCP_1970,
Paessler_pssb_1975,Huang_SS_1981,Gutsche_pssb_1982, Peuker_pssb_1982,Burt_JPC_1983,Paessler_CJP_1982,
Goguenheim_JAP_1990}. References ~\onlinecite{Huang_SS_1981,Gutsche_pssb_1982,Peuker_pssb_1982,Burt_JPC_1983},
in particular, contain important lessons about the proper application of various approaches.
Unfortunately, activity in this field stagnated, partly due to the difficulty of obtaining
sufficiently accurate results with the computational methods that were available at the time.
The current power of accurate electronic structure methods creates a fertile environment
in which to achieve the goal of a {\it quantitative} determination of carrier capture rates
completely from first principles.

This analysis, which finds its roots in similar discussions of 1970s and 1980s, relates to the underlying theory to 
describe nonradiative capture and does not undermine the achievements of Ref.~ \onlinecite{Shi_PRL_2012}. Most importantly, 
in Ref.~ \onlinecite{Shi_PRL_2012} a new efficient algorithm to calculate electron-phonon coupling was proposed.
The algorithm considers all phonon degrees of freedom and can be in principle used also in conjunction
with the static approach used in the current work. A further extension of the algorithm to hybrid functionals is 
currently highly desirable.

\subsection{Comparison with other materials\label{sum}}

It is instructive to compare our results of capture coefficients to those in other materials.
However, one should keep in mind that for any given defect the resulting capture rate
depends on many parameters, including: (i) details of the cc diagram,
for example ``classical'' barriers  for nonradiative capture $\Delta E_{\text{b}}$
(cf.~Figs.~\ref{CN-1D-CCD}, \ref{Li-1D-CCD}, \ref{ZnGa-VN-1D-CCD}); (ii) the strength
of electron-phonon coupling [Eqs. (\ref{F2}) and (\ref{r5})]; (iii) the charge state of the
defect [cf.~Eqs.~(\ref{sca1}) and (\ref{temp-scaling})]; (iv) temperature. Furthermore, many experimental
papers report capture cross sections $\sigma$ rather than capture coefficients $C$. As discussed
in Sec.\ \ref{def}, the two are related via $C=\left\langle v\right\rangle\sigma$, where $\left\langle v\right\rangle$
is a characteristic carrier velocity. In the case of non-degenerate carrier statistics
the characteristic velocity is the average thermal velocity $\left\langle v\right\rangle\sim \sqrt{T}$,
which introduces an additional temperature-dependent prefactor in the expression for $\sigma$.
These considerations indicate that the comparison of carrier capture characteristics
of different defects in different materials should be approached with caution.
However, some conclusions can still be drawn.

Let us consider a specific material, and look at carrier capture cross sections
for a wide range of defects. Those cross sections will depend on the specific defect, as well
as on temperature. However, one could argue that the maximum value of capture cross section
across all defects and across all temperature ranges (say, for temperatures where
non-degenerate carrier statistics apply), would be indicative of the strength of
electron-phonon coupling in the host material.

In their seminal paper, Henry and Lang studied capture cross sections at defects in GaP and GaAs.
\cite{Henry_PRB_1977} The measurements were performed by deep-level transient spectroscopy
over the temperature range 100-600 K, comparable to the temperature range discussed in
the present work. Both electron and hole capture was studied.
Capture cross sections at various defects ranged from $10^{-5}$ to about $100$ \AA$^2$ for
these temperatures. One could cautiously conclude that $\sim$$100$ \AA$^2$ is the maximum capture cross
section for any defect system in GaP and GaAs.

For the three defects studied in the present work, calculated hole capture cross sections vary
from 0.1 to 200 \AA$^{2}$ in GaN (Figs.~ \ref{C-nonrad} and \ref{ZnGa-VN-nonrad}), and from
1000 to 2000 \AA$^2$ in ZnO (Fig.~\ref{Li-nonrad}) for a similar temperature range.
In particular, we find $\sigma=180$ \AA$^2$ in GaN:C$_{\text{N}}$ at $T$=$600$ K, and
$\sigma=1000$ \AA$^2$ for ZnO:Li$_{\text{Zn}}$ at $T=225$ K. As discussed in Sec.~\ \ref{res}, these results are
in agreement with the experimental data of Ref.~ \onlinecite{Reshchikov_AIP_2014}. While our data set is limited, it is clear that
carrier capture cross sections in ZnO and GaN tend to be larger than those in GaP and
GaAs.  We attribute this to the larger strength of electron-phonon interactions in wide-band-gap
materials such as GaN and ZnO, compared to GaP and  GaAs.  This strength directly affects the matrix
elements, as discussed in Sec.~\ref{e-ph7}.

These trends for nonradiative capture at defects are in accord with the knowledge of
electron-phonon interactions in \emph{bulk} solids. Let us take the interaction
of free carriers with longitudinal optical (LO) phonons as an example. This strength of this interaction
can be characterized by a dimensionless factor, the Fr\"{o}hlich parameter $\alpha_{F}=e^2/\hbar\sqrt{m/(2\hbar\omega_{LO})}
(1/\varepsilon_\infty-1/\varepsilon_0)$ (in SI units).\cite{Ridley} For holes, $\alpha_{F}\approx0.15$ in GaAs 
and $\alpha_F\approx0.2$ in GaP, while $\alpha_F$ is about 1.0 in GaN and 1.5 in ZnO, indeed indicative
of stronger interaction.

\section{Conclusions \label{conc}}

In conclusion, we have developed a first-principles methodology to study nonradiative carrier capture
by means of multiphonon emission at defects in semiconductors. All the parameters,
including the electron-phonon coupling, have been determined consistently
using hybrid density functional calculations, which yield accurate bulk band structures as well as
defect properties. Significant simplifications occur due to the implementation of a 1D model, for which
we provided extensive justification, and which also yields useful insights into the defect physics.
We applied our methodology to several hole-capturing centers in GaN and ZnO. The resulting capture
coefficients are large, and in agreement with experimental data. We conclude that state-of-the-art
electronic structure techniques, when combined with reliable methodological approaches,
are capable of accurately describing carrier capture processes. The methodology
thus allows generating reliable information for defects for which experimental information is incomplete---which seems to be the case
for the majority of defects that are potentially relevant for charge trapping or SRH recombination in semiconductor devices.
The combination of a first-principles approach with experiment can also be a powerful aid in the identification of defects.
Finally, our first-principles approach allows making predictions for new materials for which no
experimental data is as yet available.


\section*{Acknowledgements}

We acknowledge M. A. Reshchikov for valuable interactions and providing us with a preprint
of Ref.~\ \onlinecite{Reshchikov_AIP_2014}.  We also thank L. Shi and L.-W. Wang for sharing
their data for GaN:(Zn$_\text{Ga}-V_{\text{N}}$), A. Carvalho for providing the
Li pseudopotential, and C. E. Dreyer, F. Giustino, A. Janotti, J. L. Lyons, J. S. Speck,
S. Keller, and C. Weisbuch for fruitful discussions.
A.A. was supported by the Office of Science of of the U.S.~Department of Energy
(Grant No.~DE-SC0010689), and by the Swiss NSF (PA00P2$\_$134127) during the initial phase of the project.
Q.Y. was supported by the UCSB Solid State Lighting and Energy Center. Computational resources were provided
by the National Energy Research Scientific Computing Center, which is supported by the DOE Office of Science
under Contract No.~DE-AC02-05CH11231, and by the Extreme Science and Engineering Discovery Environment (XSEDE),
supported by NSF (OCI-1053575 and NSF DMR07-0072N).












\begin{thebibliography}{10}
\bibitem{Mishra_IEEE_2002}
U. K. Mishra, P. Parikh, and Y.-F. Wu,
Proc. IEEE \textbf{90}, 1022 (2002).


\bibitem{Abakumov}
V. N. Abakumov, V. I. Perel', and I. N. Yassievich,
\emph{Nonradiative Recombinations in Semiconductors}
(North-Holland, Amsterdam, 1991).


\bibitem{DiBartolo}
B. Di Bartolo (ed.),
\emph{Advances in Nonradiative Processes in Solids}
(Plenum Press, New York, 1991).


\bibitem{Henry_PRB_1977}
C. H. Henry and D. V. Lang,
Phys. Rev. B \textbf{15}, 989 (1977).


\bibitem{Huang_PRCL_1950}
K. Huang and A. Rhys,
Proc. Royal Soc. \textbf{204}, 406 (1950).


\bibitem{Kubo_PTP_1955}
R. Kubo and I. Toyozawa,
Prog. Theor. Phys. \textbf{13}, 160 (1955).


\bibitem{Gummel_AP_1957}
H. Gummel and M. Lax,
Annals of Phys. \textbf{2}, 28 (1957).


\bibitem{Freed_JCP_1970}
K. F. Freed and J. Jortner,
J. Chem. Phys. \textbf{52}, 6272 (1970).


\bibitem{Paessler_pssb_1975}
R. P\"{a}ssler,
Phys. Status Solidi (b) \textbf{68}, 69 (1975).


\bibitem{Huang_SS_1981}
K. Huang,
Scientia Sinica \textbf{14}, 27 (1981).


\bibitem{Gutsche_pssb_1982}
E. Gutsche,
Phys. Status Solidi \textbf{109}, 583 (1982).


\bibitem{Peuker_pssb_1982}
K. Peuker, R. Enderlein, A. Schenk, and E. Gutsche,
Phys. Status Solidi \textbf{109}, 599 (1982).


\bibitem{Burt_JPC_1983}
M. G. Burt,
J. Phys. C: Solid State Phys. \textbf{16}, 4137 (1983).


\bibitem{Paessler_CJP_1982}
R. P\"{a}ssler,
Czech. J. Phys. \textbf{32}, 846 (1982).


\bibitem{Goguenheim_JAP_1990}
D. Goguenheim and M. Lannoo,
J. Appl. Phys. \textbf{68}, 1059 (1990).


\bibitem{Stoneham}
A. M. Stoneham, \emph{Theory of Defects in Solids}
(Oxford University Press, 1975).


\bibitem{Strehlow_JPCRD_1973}
W. H. Strehlow and E. L. Cook,
J. Phys. Chem. Ref. Data \textbf{2}, 163 (1973).


\bibitem{VanDeWalle_JAP_2004}
C. Freysoldt, B. Grabowski, T. Hickel, J. Neugebauer, G. Kresse, A. Janotti, and C. G. Van de Walle,
Rev. Mod. Phys. {\bf 86}, 253 (2014).


\bibitem{Estreicher}
{\it Theory of Defects in Semiconductors},
edited by D. A. Drabold and S. Estreicher (Springer, 2007).


\bibitem{Wiley}
{\it Advanced Calculations for Defects in Materials},
edited by A. Alkauskas, P. De\'ak, J. Neugebauer,
A. Pasquarello, and C. G. Van de Walle (Wiley, Weinheim, 2011).


\bibitem{Schanovsky_JVST_2011}
F. Schanovsky, W. G\"{o}s, and T. Grasser,
J. Vac. Sci. Technol. B \textbf{29}, 01A201 (2011).


\bibitem{Schanovsky_JCE_2012}
F. Schanovsky, O. Baumgartner, V. Sverdlov, and T. Grasser,
J. Comput. Electron. \textbf{11}, 218 (2012).


\bibitem{McKenna_PRB_2012}
K. P. McKenna and J. Blumberger,
Phys. Rev. B \textbf{86}, 245110 (2012).


\bibitem{Marcus_RMP_1993}
R. A. Marcus,
Rev. Mod. Phys. \textbf{65}, 599 (1993).


\bibitem{Shi_PRL_2012}
L. Shi and L.-W. Wang,
Phys. Rev. Lett. \textbf{109}, 245501 (2012).


\bibitem{Reshchikov_PRB_2011}
M. A. Reshchikov, A. A. Kvasov, M. F. Bishop, T. McMullen, A. Usikov, V. Soukhoveev, and V. A. Dmitriev,
Phys. Rev. B \textbf{84}, 075212 (2011).


\bibitem{Baroni_RMP_2001}
S. Baroni, S. de Gironcoli, and A. Dal Corso,
Rev. Mod. Phys. \textbf{73}, 515 (2001).


\bibitem{Giustino_PRB_2007}
F. Giustino, M. L. Cohen, and S. G. Louie,
Phys. Rev. B \textbf{76}, 165108 (2007).


\bibitem{Borrelli_JCP_2003}
R. Borrelli and A. Peluso,
J. Chem. Phys. \textbf{119}, 8437 (2003).


\bibitem{Nepal_APL_2005}
N. Nepal, J. Li, M. L. Nakarmi, J. Y. Lin, and H. X. Jiang,
Appl. Phys. Lett. \textbf{87}, 242104 (2005).


\bibitem{Reynolds_PRB_1999}
D. C. Reynolds, D. C. Look, B. Jogai, C. W. Litton, G. Cantwell, and W. C. Harsch,
Phys. Rev. B \textbf{60}, 2340 (1999)


\bibitem{Reshchikov_JAP_2005}
M. A. Reshchikov and H. Morko\c{c},
J. Appl. Phys. \textbf{97}, 061301 (2005).


\bibitem{Reshchikov_PB_2007}
M. A. Reshchikov, H. Morko\c{c}, B. Nemeth, J. Nause, J. Xie, B. Hertog, and A. Osinsky,
Physica B \textbf{401-402}, 358 (2007).


\bibitem{Reshchikov_MRS_2007}
M. A. Reshchikov, J. Garbus, G. Lopez, M. Ruchala, B. Nemeth, and J. Nause,
Mater. Res. Soc. Symp. Proc. \textbf{957}, 0957-K07-19 (2007).


\bibitem{Reshchikov_AIP_2014}
M. A. Reshchikov,
AIP Conf. Proc. \textbf{1583}, 127 (2014).


\bibitem{Ogino_JJAP_1980}
T. Ogino and M. Aoki,
Jap. J. Appl. Phys. \textbf{19}, 2395 (1980).


\bibitem{Meyer_pss_2004}
B. K. Meyer, H. Alves, D. M. Hofmann, W. Kriegsies, D. Forster, F. Bertram, J. Christen, A. Hoffmann, M. Strassburg,
M. Dworzak, U. Haboeck, and A. V. Rodina, Phys. Status Solidi B \textbf{241}, 231 (2004)


\bibitem{Lyons_APL_2009}
J. L. Lyons, A. Janotti, and C. G. Van de Walle,
Appl. Phys. Lett. \textbf{97}, 1521088 (2009).


\bibitem{Matulionis_APL_2009}
A. Matulionis, J. Liberis, I. Matulionien\.{e}, M. Ramonas, E. \v{S}ermuk\v{s}nis,
J. H. Leach, M. Wu, X. Ni, X. Li, and H. Morko\c{c},
Appl. Phys. Lett. \textbf{95}, 192102 (2009).


\bibitem{Heyd_JCP_2003}
J. Heyd, G. E. Scuseria, and M. Ernzerhof,
J. Chem. Phys. \textbf{118}, 8207 (2003);
{\it ibid.} \textbf{124}, 219906 (2006).


\bibitem{Pacchioni_PRB_2000}
G. Pacchioni, F. Frigoli, D. Ricci, and J. A. Weil,
Phys. Rev. B \textbf{63}, 054102 (2000).


\bibitem{Deak_JPCM_2005}
P. De\'{a}k, A. Gali, A. Solyom, A. Buruzs, and Th.  Frauenheim,
J. Phys.: Condens. Mat. \textbf{17}, S2141 (2005).


\bibitem{Alkauskas_PRL_2008}
A. Alkauskas, P. Broqvist, and A. Pasquarello,
Phys. Rev. Lett. \textbf{101}, 046405 (2008);
Phys. Status Solidi B \textbf{248}, 775 (2011).


\bibitem{Lyons_PRL_2012}
J. L. Lyons, A. Janotti, and C. G. Van de Walle,
Phys. Rev. Lett. \textbf{108}, 156403 (2012).


\bibitem{Lyons_JJAP_2013}
J. L. Lyons, A. Janotti, and C. G. Van de Walle,
Jap. J. Appl. Phys. \textbf{52}, 08JJ04 (2013).


\bibitem{Perdew_PRL_1996}
J. P. Perdew, K. Burke, and M. Ernzerhof,
Phys. Rev. Lett. {\bf 77}, 3865 (1996).


\bibitem{Bloechl_PRB_1994}
P. E. Bl\"{o}chl,
Phys. Rev. B \textbf{50}, 17953 (1994).


\bibitem{VASP}
G. Kresse and J. Furthm\"{u}ller, Phys. Rev. B \textbf{54}, 11169 (1996);
G. Kresse and D. Joubert, Phys. Rev. B \textbf{59}, 1758 (1999).


\bibitem{Paier_JCP_2006}
J. Paier, M. Marsman, K. Hummer, G. Kresse, I. C. Gerber, amd J. G. \'{A}ngy\'{a}n,
J. Chem. Phys. \textbf{124}, 154709 (2006).


\bibitem{Madelung}
{\it Semiconductors: Basic Data}, 2nd ed.,
edited by O. Madelung (Springer, Berlin, 1996).


\bibitem{Freysoldt_PRL_2009}
C. Freysoldt, J. Neugebauer, and C. G. Van de Walle,
Phys. Rev. Lett. \textbf{102}, 016402 (2009);
Phys. Status Sol. B \textbf{248}, 1067 (2011).


\bibitem{Baldereschi_PRB_1973}
A. Baldereschi,
Phys. Rev. B \textbf{7}, 5212 (1973).


\bibitem{Ihm_JPC_1979}
J. Ihm, A. Zunger, and M.-L. Cohen,
J. Phys. C: Solid State Phys. \textbf{12}, 4409 (1979).


\bibitem{Troullier_PRB_1991}
N. Troullier and J. L. Martins,
Phys. Rev. B {\bf 43}, 1993 (1991).


\bibitem{Fuchs_CPC_1999}
M. Fuchs and M. Scheffler,
Comput. Phys. Commun. \textbf{119}, 67 (1999).


\bibitem{CPMD}
CPMD, Copyright IBM Corp 1990-2006,
Copyright MPI f\"{u}r Festk\"{o}rperforschung Stuttgart 1997-2001;
J. Hutter and A. Curioni,
ChemPhysChem \textbf{6}, 1788 (2005).


\bibitem{Todorova_JPCB_2006}
T. Todorova, A. P. Seitsonen, J. Hutter, I. F. W. Kuo, and C. J. Mundy,
J. Phys. Chem. B \textbf{110}, 3685 (2006).


\bibitem{Broqvist_PRB_2009}
P. Broqvist, A. Alkauskas, and A. Pasquarello,
Phys. Rev. B \textbf{80}, 085114 (2009).


\bibitem{Komsa_PRB_2010}
H.-P. Komsa, P. Broqvist, and A. Pasquarello,
Phys. Rev. B \textbf{81}, 205118 (2010).


\bibitem{Alkauskas_PRL_2008b}
A. Alkauskas, P. Broqvist, F. Devynck, and A. Pasquarello,
Phys. Rev. Lett. \textbf{101}, 106802 (2008).


\bibitem{Wu_PRB_2009}
X. Wu, E. J. Walter, A. M. Rappe, R. Car, and A. Selloni,
Phys. Rev. B \textbf{80}, 115201 (2009).


\bibitem{Hedin_1970}
L. Hedin and S. Lundqvist,
in {\it Solid State Physics} (Ed.  F. Seitz, D. Turnbull and H. Ehrenreich),
Vol. 23, 1 (1970)


\bibitem{Du_PRB_2009}
M.-H. Du and S.-B. Zhang,
Phys. Rev. B \textbf{80}, 115217 (2009).


\bibitem{Carvalho_PRB_2009}
A. Carvalho, A. Alkauskas, A. Pasquarello, A. K. Tagantsev, and N. Setter,
Phys. Rev. B \textbf{80}, 195205 (2009).


\bibitem{Bjorheim_JPCC_2012}
T. S. Bj{\o}rheim, S. Erdal, K. M. Johansen, K. E. Knutsen, and T. Norby,
J. Phys. Chem. C \textbf{116}, 23764 (2012).


\bibitem{Lyons_2014}
J. L. Lyons, A. Janotti, and C. G. Van de Walle,
in preparation (2014).


\bibitem{Alkauskas_PRL_2012}
A. Alkauskas, J. L. Lyons, D. Steiauf, and C. G. Van de Walle,
Phys. Rev. Lett. \textbf{109}, 267401 (2012).


\bibitem{Bonch-Bruevich_FTT_1959}
V. L. Bonch-Bruevich,
Fiz. Tverd. Tella, Sbornik II, 182 (1959).


\bibitem{Paessler_pssb_1976}
R. P\"{a}ssler,
Phys. Status Solidi B \textbf{78}, 625 (1976).


\bibitem{Kohn_PR_1957}
W. Kohn,
Phys. Rev. \textbf{105}, 509 (1957).


\bibitem{Landau}
L. D. Landau and L. M. Lifshitz,
\emph{Quantum Mechanics. Non-relativistic theory}
(Butterworth-Heinemann, 1981).


\bibitem{Pankove_1975}
J. J. Pankove, S. Bloom, and G. Harbecke,
RCA Rev. \textbf{36}, 163 (1975).


\bibitem{Santic_SST_2003}
B. \v{S}anti\'{c},
Semicond. Sci. Technology \textbf{18}, 219 (2003)


\bibitem{Hummer_PSS_1973}
K. H\"{u}mmer,
Phys. Status Solidi (b) \textbf{56}, 249 (1973).


\bibitem{Piprek}
I. Vurgaftman and J. R. Meyer,
chapter 2 in \emph{Nitride Semiconductor Devices}, Ed.~ J. Piprek,
(Wiley-VCH, 2007).


\bibitem{Yan_SST_2011}
Q. Yan, P. Rinke, M. Winkelnkemper, A. Qteish, D. Bimberg, M. Scheffler, and C. G. Van de Walle,
Semicond. Sci. Technol. \textbf{26}, 014307 (2011).


\bibitem{Demchenko_PRL_2013}
D. O. Demchenko, I. C. Diallo, and M. A. Reshchikov,
Phys. Rev. Lett. \textbf{110}, 087404 (2013).


\bibitem{Seager_JL_2004}
C. H. Seager, D. R. Tallant, J. Yu, and W. G\"{o}tz,
J. of Lumin. \textbf{106}, 115 (2004).


\bibitem{Lyons_APL_2010}
J. L. Lyons, A. Janotti, and C. G. Van de Walle,
Appl. Phys. Lett. \textbf{97}, 152108 (2010).


\bibitem{Reshchikov_PL_2012}
M. A. Reshchikov,
in \emph{Photoluminescence: Applications, Types and Efficacy},
Ed. M. A. Case and B. C. Stout (Nova Science, New York, 2012).


\bibitem{Lyons_PRB_2014}
J. L. Lyons, A. Janotti, and C. G. Van de Walle,
Phys. Rev. B \textbf{89}, 035204 (2014).


\bibitem{Chen_PRB_2013}
W. Chen and A. Pasquarello,
Phys. Rev. B \textbf{88}, 115104 (2013).


\bibitem{Lany_PRB_2009}
S. Lany and A. Zunger,
Phys. Rev. B \textbf{80}, 085202 (2009).


\bibitem{Stoneham_RPP_1981}
A. M. Stoneham,
Rep. Prog. Phys. \textbf{44}, 1251 (1981).


\bibitem{Ridley}
See, e.g., p. 62 of B. K. Ridley,
\emph{Quantum Processes in Semiconductors}
(Clarendon Press, 1999).




\end{thebibliography}
\end{document}